\documentclass[prb,twocolumn,amssymb,amsmath,showpacs,superscriptaddress,floatfix,makecell]{revtex4-1}
\usepackage[colorlinks=true,linkcolor=blue]{hyperref}
\usepackage{graphicx,bm}
\usepackage{epstopdf}

\begin{document}

 \title{Relativistic suppression of Auger recombination in Weyl semimetals}

 \date{\today}

 \author{A.~N.~Afanasiev}
 \email{afanasiev.an@mail.ru} 
 \author{A.~A.~Greshnov}
 \affiliation{Ioffe Institute, St.Petersburg 194021, Russia}
 \affiliation{Moscow Institute of Physics and Technology, Dolgoprudny 141700, Russia}
 \author{D.~Svintsov}
 \affiliation{Moscow Institute of Physics and Technology, Dolgoprudny 141700, Russia}

\begin{abstract}
Auger recombination (AR) being electron-hole annihilation with energy-momentum transfer to another carrier is believed to speed up in materials with small band gap. We theoretically show that this rule is violated in gapless three-dimensional materials with ultra-relativistic electron-hole dispersion, Weyl semimetals (WSM). Namely, AR is prohibited by energy-momentum conservation laws in {\it prototypical} WSM with a single Weyl node, even in the presence of anisotropy and tilt. In {\it real} multi-node WSM, the geometric dissimilarity of nodal dispersions enables weak {\it inter-node} AR, which is further suppressed by strong screening due to large number of nodes. While partial AR rates between the nodes of the same node group are mutually equal, the inter-group processes are non-reciprocal, so that one of groups is {\it geometrically protected} from AR. This geometrical protection prolongs AR lifetime up to two orders of magnitude, to the level of nanoseconds.
\end{abstract}

\maketitle

%%%%%%%%%%%%%%%%%%%%%%%%%%%%%%%%%%%%%%%%%%%%%%%%%%%%%%%%%%%%%%%%%%%%%%%
%%%%%%%%%%%%%%%%%%%%%%%%%%%%%%%%%%%%%%%%%%%%%%%%%%%%%%%%%%%%%%%%%%%%%%%
%\section{Introduction\label{sec1}}
{\it Introduction.---} The latest years of condensed matter physics are marked by an intense search for solid-state realizations of exotic fundamental particles~\cite{Armitage2018,volovik2003,Elliott_Majorana}. The unique electronic properties of graphene~\cite{Novoselov2005}, Weyl~\cite{Lv2015TaAs,Xu2015TaAs,Xu2015TaP,Xu2015NbAs,Devizorova_Weyl}, and Dirac~\cite{Liu_Dirac_Na3Bi} semimetals enabled the tabletop observation of Klein tunneling~\cite{Stander_KleinGraphene}, supercritical atomic collapse~\cite{Levitov_AtomicCollapse}, axial~\cite{zhang_ABJAnomaly} and axial-gravitational~\cite{Gooth2017} anomalies. In this strive for high-energy physics {\it enabled} by electronic properties of novel materials, less attention is paid to the effects {\it prohibited} by these properties. Such negatory search still can be fruitful. In particular, suppression of electron scattering, relaxation, and recombination in solids~\cite{BackscatteringSuppressionWeyl,BakcscatteringSuppressionGraphene,BackscatteringKibis} would enable the observation of new phases of ultra-clean matter, not to say about ultrafast electronic and photonic devices.

In this Letter, we show that quasi-relativistic dispersion of fermions in recently discovered Weyl semimetals (WSM) strongly suppresses the electron-hole recombination with energy-momentum transfer to another carrier, known as Auger recombination (AR). The AR is among key obstacles toward the realization of non-equilibrium phases of electron-hole liquid~\cite{keldysh1986EHL}, excitonic~\cite{Triola_Insulator} and Floquet topological insulators~\cite{Floquet_Insulator}. In addition, AR is the primary ''killer'' of population inversion and optical gain in narrow-gap semiconductor lasers~\cite{Iveland_AugerDroop,Morozov_StimulatedHgTe} hindering their promotion into terahertz range. Numerous attempts to suppress AR invoked strain-engineering~\cite{Adams_strain}, modification of wave function profiles~\cite{Efros_Suppression}, and exchange effects upon scattering~\cite{Magnetic_suppression}. However, there has been no material with ''natural'' AR suppression as it occurs in WSM.

\begin{figure}[t]
 \includegraphics[width=86mm]{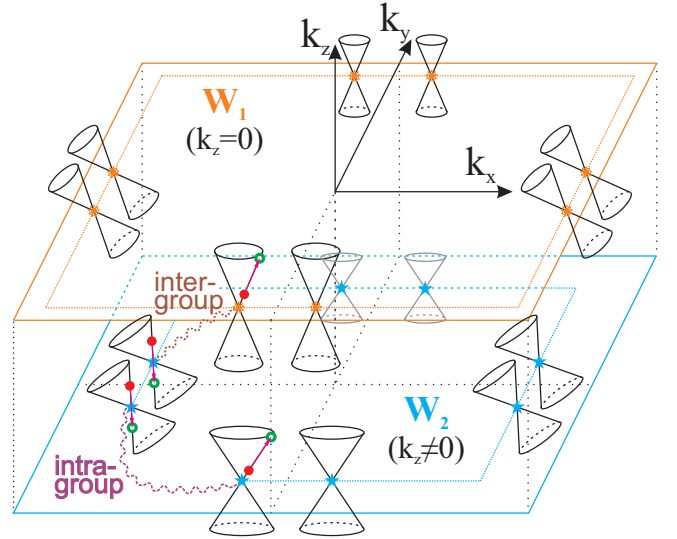}
 \caption{\label{Fig:Nodes}
Schematic structure of the Weyl nodes and inter-node AR in TaAs family of WSM having $C_{4v}$ point symmetry with two groups of the nodes, known as $W_1$ and $W_2$. %Both $W_1$ and $W_2$ are submitted to $C_{4v}$ point group but $W_1$ contains only half number (8) of the nodes of $W_2$ (16) since it lies on the symmetry plane $k_z=0$, while upper plane of $W_2$ duplicates the lower one and is omitted here.
Wavy lines show the two types of the Auger processes involving two nodes of the same group or the nodes of different groups. In the latter case, the interband transition occurs in $W_1$ and the intraband one -- in $W_2$ (or vice versa).}
\end{figure}

We further elaborate the incompleteness of analogy between relativistic electrons and Weyl fermions, and reveal its consequences for AR. In real WSM, there exist multiple Weyl nodes located at {\it low-symmetry} points of the Brillouin zone (Fig.~\ref{Fig:Nodes}), which can be attributed to one or several node groups ($W_1$ and $W_2$ in TaAs family). As a result, the carrier energy spectrum acquires anisotropy~\cite{Arnold2016,Hu2016,Klotz2016} and tilt. We show that it is the dissimilarity of energy spectra at different nodes that enables weak {\it inter-node}  recombination, while the {\it intra-node} AR remains prohibited. Similar inter-valley AR was noted for lead salts~\cite{Emtage_AR_Lead_tin}, however, strong suppression of intra-valley process was not realized. Adopting the relation between recombination and dissipative part of polarizability~\cite{Ziep1980}, we derive an illustrative geometric construction for evaluation of AR. We use it to reveal the key features of interband population inversion dynamics. The most intriguing one is non-reciprocity of AR between various node groups, and suppression of AR at nodes with fast carrier velocity, which we call {\it geometrically protected population inversion}. This protection, along with strong screening due to large number of nodes, prolongs the non-radiative lifetimes up to the level of several nanoseconds.
%We factorize the expression for AR rate (ARR) into ''geometrical part'' characterizing the dissimilarity in carrier dispersion of the nodes, and ''statistical part'' characterizing the band occupation. Using the obtained expression, we study the relaxation dynamics of interband population inversion in various regimes and show that it can be {\it sub-exponential}, {\it super-exponential}, and {\it multi-exponential}.

%%%%%%%%%%%%%%%%%%%%%%%%%%%%%%%%%%%%%%%%%%%%%%%%%%%%%%%%%%%%%%%%%%%%%%%
%%%%%%%%%%%%%%%%%%%%%%%%%%%%%%%%%%%%%%%%%%%%%%%%%%%%%%%%%%%%%%%%%%%%%%%
{\it AR in prototypical WSM.---} 
The suppression of AR in WSM is tightly linked to the impossibility of impact ionization of Dirac vacuum by high-energy electrons~\cite{Dirac1930}. Indeed, an electron with energy-momentum relation $E^2 = (mv_0^2)^2+(kv_0)^2$ cannot emit electron-positron pairs due to non-equal energies of initial ($mv_0^2$) and final states ($\ge 3mv_0^2$) in the center-of-mass system. 

As the mass gap tends to zero and dispersion becomes {\it ultra-relativistic}, $E({\bf k})= kv_0$, the situation becomes pathological. Momentum conservation for Auger process reads ${\bf k}_{e1}+{\bf k}_{h1} + {\bf k}_{e2} = {\bf k}'_{e2}$, where ${\bf k}_{e1}$ and ${\bf k}_{h1}$ are the momenta of recombining electron and hole (counterpart of positron in solids), ${\bf k}_{e2}$ and ${\bf k}'_{e2}$ are the initial and final momenta of 'hot' electron. The energy conservation implies ${k}_{e1}+{k}_{h1} + {k}_{e2} = {k}'_{e2}$, and renders all four momenta collinear. The phase space for collinear collisions vanishes. However, the interaction strength between collinear carriers diverges as their unidirectional motion with equal velocities implies infinite interaction time. This fact is known as collinear scattering anomaly~\cite{Sachdev_QuantumCritical}, and the resulting AR probability of the form $0\times \infty$ was shown to be finite in two dimensions~\cite{Rana_Auger_2007,Malic-Auger,Tomadin-theory,Svintsov2018}. 

To put the solution of AR problem in three dimensions on a solid ground, we use a transparent yet not widely adopted relation~\cite{Ziep1980} between recombination and imaginary parts of inter- and intraband polarizations, ${\rm Im}\Pi_{-+}$ and ${\rm Im}\Pi_{ss}$ (here $s=\pm 1$ is the index of conduction and valence bands). The AR rate in this formalism involves the product of electron-hole annihilation probability characterized by ${\rm Im}\Pi_{-+}$ , the squared amplitude of virtual photon propagation, and the probability of interband photon absorption ${\rm Im}\Pi_{ss}$. The formal expression for the rate of AR with energy transferred to the $s$-th band, ${\mathcal R}^{(s)}$, reads
\begin{multline}
 \label{AR-rate}
 {\mathcal R}^{(s)} = 4 \sum\limits_{{\bf q}\omega}  {\rm Im}\Pi_{-+}(\omega,{\bf q})
 \frac{|V_0(q)|^2}{|\epsilon(\omega,{\bf q})|^2} {\rm Im}\Pi_{ss}(\omega,{\bf q}) \times \\
 \left[n_B(\omega - \Delta\mu_{eh}) - n_B(\omega)\right],
\end{multline}
where $V_0({\bf q})=4\pi e^2/q^2$ is the Fourier transform of Coulomb interaction, $\epsilon(\omega,{\bf q})$ is the dielectric function of WSM, $n_B(\omega) = [e^{\omega/T}-1]^{-1}$ is the Bose distribution, and $\Delta\mu_{eh} = \mu_e - \mu_h$ is the difference of electron and hole quasi-Fermi levels~\footnote{We note that Eq.~(\ref{AR-rate}) incorporates both {A}uger recombination and generation processes, therefore it vanishes in equilibrium.}. Summation in (\ref{AR-rate}) is performed over all possible frequencies $\omega$ and wave vectors ${\bf q}$ of virtual photons, $\sum_{{\bf q}\omega}\equiv (2\pi)^{-4}\int{d^3{\bf q}d\omega}$. 

\begin{figure}[t]
 \includegraphics[width=0.9\linewidth]{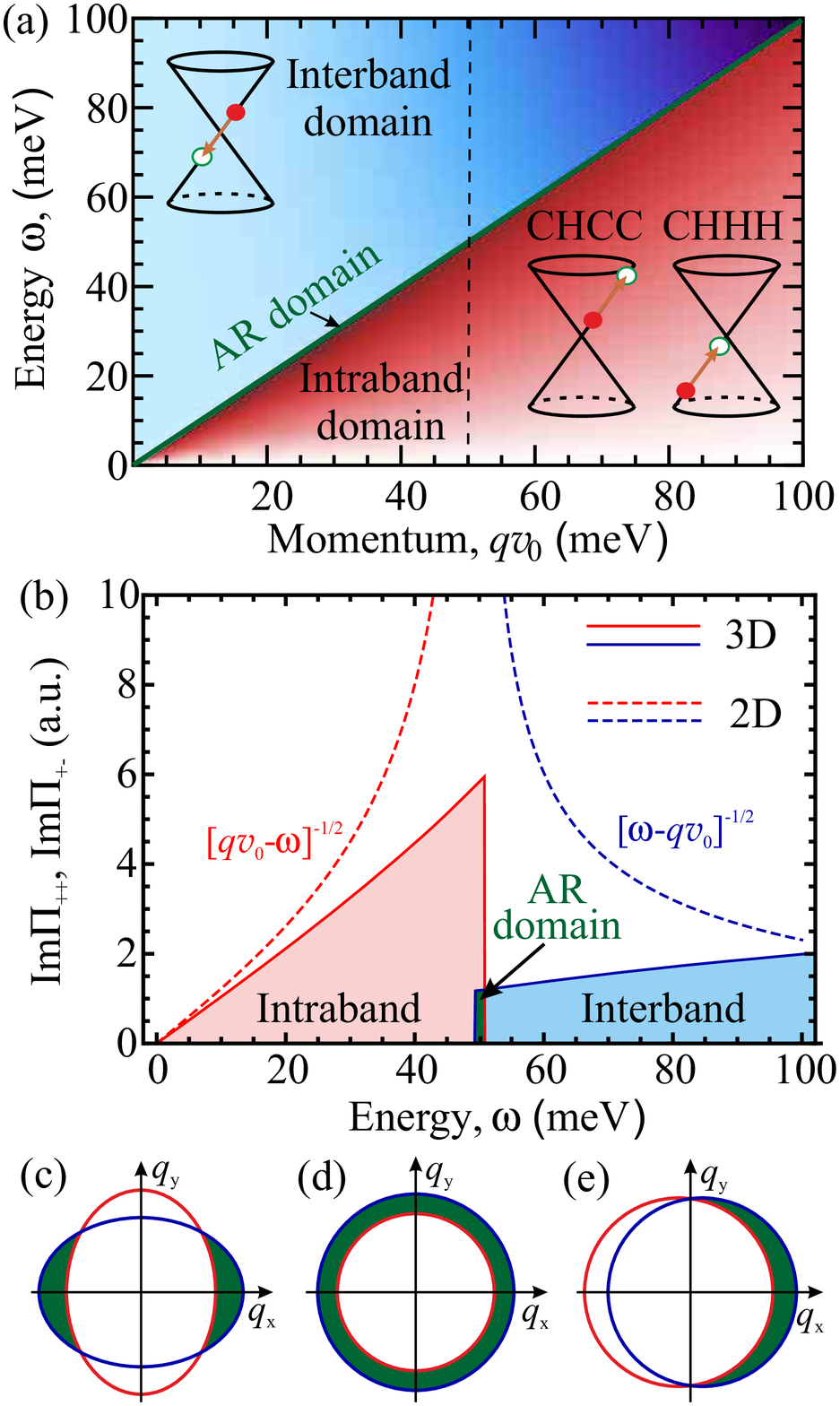}
 \caption{\label{Fig:Overlap}
Strong suppression of AR in WSM explained in terms of overlap between intra- and interband excitations (a) Color maps of inter- ($\rm{Im}\Pi_{-+}$) and intraband ($\rm{Im}\Pi_{ss}$) polarizabilities that are non-zero on different sides of $\omega=v_0q$ line in prototypical WSM. (b) Frequency dependence of polarizations at $qv_0=50$ meV. Dashed lines show the respective polarizations for 2d graphene (c-e) Overlap of the interband and intraband domains of different nodes in WSM in the $\bf q$-space at fixed frequency $\omega$ ($q_z$ is omitted for simplicity) in the cases of anisotropy (c), velocity difference (d) and tilt (e).}
\end{figure}

Since AR involves both inter- and intraband electron transitions, the domains of non-zero ${\rm Im}\Pi_{-+}$ and ${\rm Im}\Pi_{ss}$ should intersect in $(\omega, {\bf q})$ 4-space. However, the interband emission is bound inside the ''excitation cone'', $\omega \ge v_0q$, while intraband absorption is possible only outside of it, $\omega \le v_0q$, as follows from conservation laws for single-particle processes [Fig.~\ref{Fig:Overlap}(a)]. In a formal language:
\begin{gather}
\label{P_inter}
{\rm Im}\Pi^{(0)}_{-+} = F_{-+}(\omega,q)\theta(\omega - v_0q),\\
\label{P_intra}
{\rm Im}\Pi^{(0)}_{ss} = F_{ss}(\omega,q)\theta(v_0q - |\omega|),
\end{gather}
where $\theta(x)$ is the unit step function and $F_{ss'}(\omega,q)$ are smooth~\cite{SupportingInfo}.
%(the expressions are given in Supplemental Section I~\cite{SupportingInfo}).
Integration of smooth functions in (\ref{AR-rate}) over the region of zero measure results in zero value of ${\mathcal R}^{(s)}$ (see Fig.~\ref{Fig:Overlap}(b)). It contrasts to the two-dimensional case (upper curves in Fig.~\ref{Fig:Overlap}(b)) where $F_{ss'}(\omega,q)$ possess square-root singularities at $\omega = v_0q$, which resulted in finite AR rate~\cite{Rana_Auger_2007, Tomadin-theory}.

{\it AR in real WSM.---} To study AR in materials with general anisotropic Weyl velocity tensor ${\hat v}_n$ and tilt velocity  ${\bf u}_n$ (which are specific to the Weyl node group, $W_n$), we introduce the Hamiltonian 
%The impossibility of simultaneous inter- and intraband transitions may look fragile against the deformations of initially isotropic Weyl spectrum, $E = \pm k v_0$. To study AR in real materials, we introduce the Weyl velocity tensor ${\hat v}_n$ and tilt velocity  ${\bf u}_n$, which are specific to the Weyl node group, $W_n$. Generic Hamiltonian around one of the equivalent nodes within the group, ${\bf K}_n$, reads
\begin{equation}
 H_n({\bf k}) = {\bf w}_n({\bf k}){\bm \sigma} +  w^{(t)}_n({\bf k}) \sigma_0,
 \label{eq6}
\end{equation}
where ${\bf w}_n({\bf k}) = {\hat v}_n {\bf k}$ and $w_n^{(t)}({\bf k}) = {\bf u}_n {\bf k}$. Its eigenvalues are
\begin{equation}
 E_{n,s}({\bf k}) =  s|{\bf w}_n({\bf k})| +  w^{(t)}_n({\bf k}).
 \label{eq7}
\end{equation}
%where $s$ is the band index ($+1$ for electrons and $-1$ for holes).
Explicit form of dispersion in the nodes of the group are given by
%The dispersions are replicated to other nodes of the group using relation 
$E_{n,s}^{(i)}({\bf k}) = E_{n,s}(g^{(i)}_n{\bf k})$, where $g^{(i)}_n$ represent operations of the point group connecting the nodes ($C_{4v}$ for TaAs family, Fig.~\ref{Fig:Nodes}).

Linearity of the Hamiltonian (\ref{eq6}) in momentum operators allows one to relate the polarization of anisotropic tilted $i$-th node $\Pi_{ss'}^{(i)}$ and the polarization of prototypical isotropic WSM $\Pi_{ss'}^{(0)}$, via a linear coordinate transform
\begin{equation}
 \Pi_{ss'}^{(i)}(\omega,{\bf q}) = \Pi^{(0)}_{ss'}(\omega-{\bf u}_n g^{(i)}_n {\bf q}, v_n^{-1} \hat{v}_n g^{(i)}_n{\bf q}),
 \label{P_intra2}
\end{equation}
where $v_n=|\det{\hat v}_n|^{1/3}$ is the average Weyl velocity. This transform implies that anisotropy and tilt of dispersion translates in the respective deformation of the ''excitation cone''. As a result, the inter- and intra-band excitations within the same node still do not overlap in the presence of tilt and anisotropy, and intra-node AR remains forbidden. 

At the same time, deformation of the Weyl cones opens up the inter-node recombination channel. Geometrically, the overlap between inter- and intraband excitations in different nodes can be produced both by different orientation of velocity tensors, tilt, and different absolute values of Weyl velocity (in multi-group WSM). These possibilities are illustrated in Fig.~\ref{Fig:Overlap}(c-e).
%Different orientation of the velocity tensors at the nodes produces overlap between interband and intraband excitations of these nodes [Fig.~\ref{Fig:Overlap}(c)]. In multi-group case, the overlap appears due to different absolute values of Weyl velocity in various groups, even if velocity is isotropic [Fig.~\ref{Fig:Overlap}(d)]. For both cases, finite overlap is also produced by tilt [Fig.~\ref{Fig:Overlap}(e)].
As far as difference in dispersion of the nodes is small, the {\it partial} AR rate involving the nodes $i$ and $j$, $R^{(s)}_{ij}$, can be factorized  into the {\em geometric} and {\em statistical} parts~\cite{SupportingInfo},
\begin{equation}
 R^{(s)}_{ij}=\mathcal{G}^{(s)}_{ij} \mathcal{S}_{ij}^{(s)}.
 \label{eq2}
\end{equation}
%as proved in Supplemental Section II~\cite{SupportingInfo}.

The dimensionless geometric factor $\mathcal{G}$ originates from integration of the Coulomb amplitude $V_0({\bf q})$ over the allowed wave vectors [shaded areas in Fig.~2(c-e)]. Mathematically, it is given by an integral over the solid angle of a unit vector ${\bf e_q}$,
\begin{gather}
 \mathcal{G}^{(s)}_{ij} = 
 \int\limits_{\Delta({\bf e_q})>0}  \frac{\Delta({\bf e_q})d{\bf e_q}}{|v_n[g_n^{(i)}]^{-1}{\hat v}_n^{-1}{\bf e_q}|^4}
 \label{eq9}, \\
 \Delta({\bf e_q}) = 
 \left|{\hat v}_{n'}g_{ij}{\hat v}_{n}^{-1}{\bf e_q}\right|-1-({\bf u}_n-{\bf u}_{n'}g_{ij}){\hat v}_n^{-1}{\bf e_q},
 \label{eq10}
\end{gather}
where $g_{ij}=g_{n'}^{(j)}[g_{n}^{(i)}]^{-1}$, and the nodes $i,j$ are assumed to belong to the groups $W_{n,n'}$, respectively.
%, and $\Delta$ is the ''thickness'' of layer of allowed frequencies $\omega$ normalized by $v_0 q$. 

%The $\mathcal{S}$-factor is governed by the distribution of equilibrium and non-equilibrium carriers.
Once a small difference in node dispersions is carried to $\mathcal G$, the statistical factor $\mathcal S$ is expressed via the polarization of the prototypical WSM,
\begin{equation}
 \mathcal{S}^{(s)}_{ij} = \int \frac{\omega_{\bf q}d^3{\bf q}}{16\pi^5} |{\tilde V}(q)|^2
 F_{-+}^{(i)}(\omega_{\bf q},q) F^{(j)}_{ss}(\omega_{\bf q},q) \mathcal{N}_B(\omega_{\bf q}),
\end{equation}
where ${\tilde V}(q)=V_0(q)/\epsilon(\omega_{\bf q},q)$, $\mathcal{N}_B(\omega)=n_B(\omega-\Delta\mu_{eh})-n_B(\omega)$, and the frequency of virtual photon is at the edge of ''excitation cone'', $\omega_{\bf q}={v_0}q$. The factorization allows one to evaluate and discuss the effect of a particular band structure and that of statistics independently.
%, and the upper index of $F$ is due to inequality in Fermi level position relative to the Weyl points at the nodes of different groups.

We first discuss the geometry of intra-group AR enabled by in-plane velocity anisotropy, assuming principal axis of the Weyl velocity tensor parallel to the crystallographic ones. In the $W_1$ node group of TaAs, there exist 
%operations of point group $C_{4v}$ generate 
four Weyl velocity tensors with pairwise perpendicular dispersion surfaces.
%with perpendicular orientation of the constant energy surfaces 
%at the eight nodes of $W_1$ shown in Fig.~\ref{Fig:Nodes},
%\begin{equation}
% \hat{v}^{(i)}={\rm diag}(v_x\cos{\varphi_i}-v_y\sin{\varphi_i}, v_y\cos{\varphi_i}+v_x\sin{\varphi_i}, v_z),
%\end{equation}
%where $\varphi_i=0,\pi/2,\pi,3\pi/2$.
%\begin{equation}
% \hat{v}^{(i)}={\rm diag}(\pm v_y, \pm v_x, v_z).
%\end{equation}
%There are four nodes ''$j$'' with dispersion perpendicular to that of a given node ''$i$'', and each pair gives equal contribution to the total ARR, which
For each such pair, the geometric part ${\mathcal G}_{xy}$ is given by volume between two ellipsoids in ${\bf q}$-space, as illustrated in Fig.~\ref{Fig:Overlap}(c). Explicit calculation in the limit $|v_x-v_y|\ll v_{\perp}=(v_x+v_y)/2$
%(see Supplemental Section III~\cite{SupportingInfo})
results in~\cite{SupportingInfo}
\begin{align}
 \mathcal{G}_{xy} = \frac{|v_x-v_y|}{v_{\perp}} g(1-v_z^2/v_\perp^2)\left(\frac{v_z}{v_{\perp}}\right)^{5/3}, \\
 g(x)=\frac{\sqrt{x(1-x)}+(2x-1)\arctan{\sqrt{\frac{x}{1-x}}}}{(1-x)x^{3/2}}.
\end{align}

The AR rate does also not vanish even if node dispersions differ in tilt only. The tilt-enabled geometric factor $\mathcal{G}_{t}$ is given by the volume between two shifted spheres in ${\bf q}$-space, as shown in Fig.~\ref{Fig:Overlap}(e), so that
\begin{equation}
 \mathcal{G}_{t}=\frac{\pi|{\bf u}^{(i)}-{\bf u}^{(j)}|}{v_0},
\end{equation}
where ${\bf u}^{(i,j)}={\bf u}_{n,n'}g_{n,n'}^{(i,j)}$, $i,j \in W_{n,n'}$, and we have assumed equal velocity tensors for different nodes.

When there exist multiple Weyl node groups, the difference in (averaged) Weyl velocities becomes a more important factor enabling AR than anisotropy and tilt. Neglecting the latter, the geometry term ${\mathcal G}_{\circledcirc}$ is given by volume between two spheres in $q$-space, as shown in Fig.~\ref{Fig:Overlap}(d), and results in  
\begin{equation}
 \label{eqGcircledcirc}
 \mathcal{G}_{\circledcirc} = \frac{8\pi(v_2-v_1)\theta(v_2-v_1)}{v_1+v_2}.
\end{equation}
A striking feature of inter-node AR is its {\it non-reciprocity}, i.e. the absence of recombination in the group with fastest Weyl velocity $v_\star$. Indeed, the virtual photon with momentum $q<\omega/v_\star$ emitted upon e-h annihilation in ''fast'' group cannot be absorbed in ''slow'' nodes (while the opposite is possible). Thus, this peculiar ''fast'' node is {\it geometrically protected} from AR, provided the anisotropy and tilt are weak. Below we show that geometrical protection elongates the non-radiative lifetime in real WSM up to several nanoseconds.

%Though distortions of initially isotropic Weyl cones open up inter-node recombination, there are several specific non-equilibrium distributions remaining immune to AR. The simplest example of such {\it geometrically-protected} population inversion can be formed in WSM with two groups of isotropic nodes and dissimilar velocities $v_2>v_1$ under selective pumping of ''fast'' group. In this case, the momentum of emitted virtual photon $q<\omega/v_2$ is insufficient to induce intra-band absorption in the ''slow'' node. Selective pumping of groups can be based on Pauli blocking of pump photons due to dissimilar group doping.

%%%%%%%%%%%%%%%%%%%%%%%%%%%%%%%%%%%%%%%%%%%%%%%%%%%%%%%%%%%%%%%%%%%%%%%
%%%%%%%%%%%%%%%%%%%%%%%%%%%%%%%%%%%%%%%%%%%%%%%%%%%%%%%%%%%%%%%%%%%%%%%
{\it Visualization of AR in WSM: temporal dynamics.---}
Once the interband population inversion is created in WSM, its relaxation after relatively fast intraband equilibration (due to electron-electron scattering) is governed by the interband processes. Assuming that AR lifetime in WSM is shorter than the phonon-limited lifetime~\cite{Huang_2018_HotCarrier}, we consider the temporal dynamics with AR channel only,  
\begin{equation}
 \frac{d p_n}{d t} = -\mathcal{R}_n,
 \label{Eq:dyn1}
\end{equation}
where ${\mathcal{R}}_n=\sum_{\substack{i \in n\\ j,s}} \mathcal{R}^{(s)}_{ij}$
is a partial AR rate in $W_n$ and $p_n(t)=n_n(t)$ are the non-equilibrium carrier densities at each node group $W_n$. Assuming uniform pumping at each node, we solve Eq.~(\ref{Eq:dyn1}) with a given initial total non-equilibrium density $p_0$ distributed among the node groups according to their degeneracy factors $\eta_n$. For estimates, we take the average Weyl velocity  ${v_0}=2.5\cdot10^7~\rm{cm/s}$, $\varkappa=10$, $T=77~\rm{K}$, and the geometry factors $\mathcal{G}_{1,2}$ for the intra-group and inter-group contributions as 0.1 and 0.2, respectively. The dielectric function $\epsilon(\omega,{\bf q})$ is adopted in Thomas-Fermi approximation~\cite{SupportingInfo}. %The latter is proportional to the total number of nodes $\eta=\sum_n \eta_n$ (24 in TaAs family),  as well as each polarization $\Pi$ in Eq.~(\ref{AR-rate}). This makes the total AR rate (almost) independent of the nodal degeneracy and the lifetimes -- nearly proportional to $\eta$, thus enhancing them by more than an order of magnitude. 

%where $p_n(t)=n_n(t)$ are the partial densities of non-equilibrium carriers  at each node group $W_n$, and ${\mathcal{R}_n=\sum_{\substack{i \in n\\ j,s}} \mathcal{R}^{(s)}_{ij}}$ is a partial AR rate governing carrier density in $W_n$. Assuming uniform pumping at each node, we solve Eq.~(\ref{Eq:dyn1}) with a given initial total non-equilibrium carrier density $p_0$ distributed among the groups according to degeneracy factors $\eta_n$.

%To obtain realistic values of AR rate, we use Thomas-Fermi approximation for the screening function $\epsilon(\omega,{\bf q})$. The latter is proportional to the total number of nodes $\eta=\sum_n \eta_n$, equal to 24 in TaAs family (see Supplemental Material Sec.IV). Since each polarization $\Pi$ in Eq.~(\ref{AR-rate}) is proportional to $\eta$ as well, the total AR rate is (almost) independent of it, and the lifetimes are (nearly) proportional to $\eta$, which makes the lifetimes in multi-node WSM much longer. For estimates, we take $v_0=2.5\cdot10^7~\rm{cm/s}$, $\varkappa=10$, $T=77~\rm{K}$, and the geometry factors $\mathcal{G}_{1,2}$ for the intra- and inter-group contributions as 0.1 and 0.2, respectively.

\begin{figure}[t]
  \includegraphics[width=86mm]{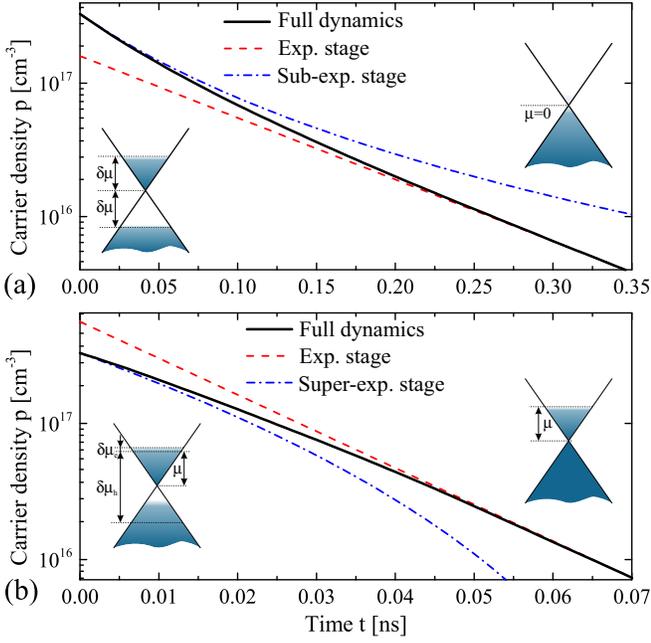}
 \caption{\label{Fig:Dyn_intra}
Dynamics of the non-equilibrium carrier density $p(t)$ in WSM with a sole node group ($p_0=3\cdot10^{17}~\rm{cm^{-3}}$), intrinsic (a) and extrinsic (b) with residual density $n_r=10^{18}~\rm{cm^{-3}}$ ($\mu\approx 25\,\rm{meV}$). The exponential (non-degenerate) and non-exponential (degenerate) analytical limits are shown by dashed lines. Insets show the band fillings right after intraband thermalization (left) and long after characteristic recombination time (right)}
\end{figure}

\begin{figure}[t]
 \includegraphics[width=86mm]{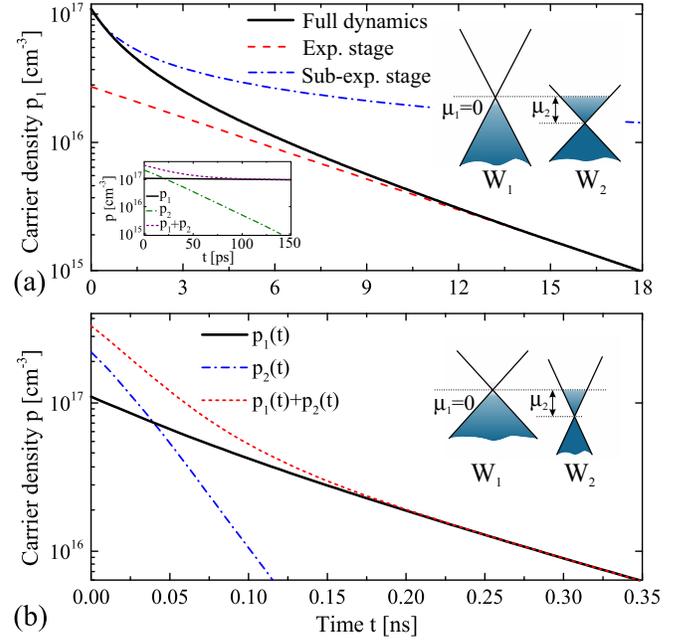}
 \caption{\label{Fig:Dyn_inter}
Dynamics of the population inversion in WSM with 8-fold degenerate intrinsic node group $W_1$ and 16-fold degenerate extrinsic group $W_2$ with (a) and without (b) {\em geometrical protection} of $W_1$ (either $v_1>v_2$ or $v_1<v_2$). The inset in (a) shows fast relaxation of $p_2$ in contrast to long-living $p_1(t)$. Diagrams on the right show the band occupancy at equilibrium}
\end{figure}

%As a first example, we consider WSM with a single group of eight nodes, as expected in strained HgTe~\cite{Ruan2016HgTe}, and the results are summarized in Table~\ref{Tab:DynamicsSummary} and Fig.~\ref{Fig:Dyn_intra}.
The specific band structure of WSM imprints on AR rates already in materials with single node group (as expected in strained HgTe~\cite{Ruan2016HgTe}). Particularly, the rate at strong pumping (quasi-Fermi level $\delta\mu \gg T$) never follows the $p^3$-law inherent to common semiconductors. The superseding dynamics is summarized in Table~\ref{Tab:DynamicsSummary} and Fig.~\ref{Fig:Dyn_intra}. 

Strong deviations from $p^3$-law occur due to carrier degeneracy and screening. The latter is proportional to the total number of nodes $\eta=\sum_n \eta_n$ (24 in TaAs family and 8 in strained HgTe). This makes the total AR rate (almost) independent of the nodal degeneracy and the lifetimes -- nearly proportional to $\eta$, thus enhancing them by more than an order of magnitude. Strong pumping of intrinsic nodes (equilibrium Fermi level $\mu=0$) results in sub-exponential relaxation, while strong pumping of extrinsic ones ($\mu\gg T$) -- in super-exponential one. Recombination at weak pumping and/or at final stages is, as expected, exponential. It is faster in extrinsic nodes by almost an order of magnitude due to the {\it parabolic} density of states in WSM.

\begin{table}[hb]
\begin{tabular}{|c|c|c|}
\hline
                                                                      & \textbf{\begin{tabular}[c]{@{}c@{}}Strong pumping\\ $\Delta\mu_{eh}\gg T$\end{tabular}}                                                                                                   & \textbf{\begin{tabular}[c]{@{}c@{}}Weak pumping\\ $\Delta\mu_{eh}\ll T$\end{tabular}}                                      \\ \hline
\textbf{\begin{tabular}[c]{@{}c@{}}Intrinsic\\ $\mu=0$\end{tabular}}  & \begin{tabular}[c]{@{}c@{}}Sub-exponential\\ $p/p_0 = (1+t/3\tau_0)^{-3}$\\ $\tau_0 = \dfrac{\eta^{4/3}}{C_{1d}(\alpha_{\eta}) \mathcal{G}_1 {v_0} p_0^{1/3}}$\end{tabular}          & \begin{tabular}[c]{@{}c@{}}Exponential\\ $\tau =\dfrac{\eta}{C_{1n}(\alpha_{\eta})\mathcal{G}_1 T} $\end{tabular}     \\ \hline
\textbf{\begin{tabular}[c]{@{}c@{}}Extrinsic\\ $\mu\gg T$\end{tabular}} & \begin{tabular}[c]{@{}c@{}}Super-exponential\\ $p/p_0 = (1-t / 3\tau_{0})^3$\\ $\tau_{0} = \dfrac{v_0 {\eta}^{2/3}p_0^{1/3}}{C_{2d}(\alpha_{\eta}) \mathcal{G}_1\mu^2}$\end{tabular} & \begin{tabular}[c]{@{}c@{}}Exponential\\ $\tau=\dfrac{\eta T}{C_{2n}(\alpha_{\eta})\mathcal{G}_1 \mu^2}$\end{tabular} \\ \hline
\end{tabular}
 \caption{\label{Tab:DynamicsSummary}
Summary of results for relaxation dynamics of excited carrier density $p(t)$ in WSM with single node group. $\mu$ is the equilibrium Fermi energy, $C(\alpha_\eta)$ are dimensionless screening functions depending on coupling constant $\alpha_{\eta}=\eta e^2/\varkappa {v_0}$ listed in~\cite{SupportingInfo}.
%Supporting section IV~\cite{SupportingInfo}
}
\end{table}

In multi-group WSM the picture of relaxation becomes much more complicated~\cite{SupportingInfo}.
%The recombination dynamics in multi-group WSM even at weak pumping is {\it multi-exponential} due to the presence of inter-group recombination that mixes carrier densities of individual groups, $p_1$ and $p_2$~\cite{SupportingInfo}.
The most remarkable feature of it %multi-group  relaxation dynamics
is the {\it geometric protection} of population inversion in ''fast'' nodes. If all eigenvalues of the Weyl velocity tensor in $W_1$ exceed those in $W_2$, the inter-group channel of AR for $W_1$ is locked. Additionally, the intra-group channel is suppressed due to screening by the resident carriers at $W_2$. The band structure of TaAs, the first experimentally discovered WSM, favours such scenario. It possesses almost intrinsic 8-fold degenerate node group $W_1$, and extrinsic 16-fold degenerate node group $W_2$ with $\mu\approx 20\,\rm{meV}$~\cite{Armitage2018}. Therefore, one can expect extra-long lifetimes of carriers in $W_1$, which is further proved by our calculations in Fig. 4(a). 

At the excitation density $p_0=3\cdot 10^{17}\,\textrm{cm}^{-3}$ the distributions at both $W_1$ and $W_2$ are degenerate, and the evolution of $p_2(t)$ is given by equations of super-exponential case with $\eta$ substituted by degeneracy of heavily occupied node $\eta_2$, while
\begin{gather}
 \label{Eq:dyn_mult_sub}
 p_1(t) = \frac{{p_1}(0)}{(1+5t/3\tau_{0})^{3/5}}, \\
 \label{Eq:tau_mult_sub}
 \tau^{-1}_{0} = C_{3n}v_0^5\eta_2^{-2}\eta_1^{-2/3}\mathcal{G}_1 \mu^{-4} p_1(0)^{5/3}
\end{gather}
with $C_{3n}\approx 234$.
%(Supplemental Section IV~\cite{SupportingInfo}).
%We emphasize that~(\ref{Eq:dyn_mult_sub},\ref{Eq:tau_mult_sub}) significantly differs from~(\ref{Eq:dyn_sub}),\ref{Eq:tau_sub}) due to the strong screening by carries occupying $W_2$ node group, which is the additional source of relaxation elongation at $W_1$.
In the non-degenerate limit of carrier distribution in $W_1$ and the valence band of $W_2$, the relaxation is exponential in both groups. The lifetime of carriers in $W_2$ is the same as in doped single-group case with $\eta\rightarrow\eta_2$, while for $W_1$
%In the weak excitation regime carrier distributions in $W_1$ and the valence band of $W_2$ are non-degenerate, so the relaxation is exponential in both node groups, but with different lifetimes given by Eq.~(\ref{Eq:tau2n}) with $\eta$ substituted by $\eta_2$ for $W_2$ and by
\begin{equation}
 \tau^{-1}=C_{1n}(\beta_{\eta_2}) \mathcal{G}_1 \eta_1 {\eta_2}^{-2} \alpha_{\eta_1}^2 \beta_{\eta_2}^{-2}T,
\end{equation}
where $\beta_{\eta_2}=(3/\pi^2)(\mu/T)^2\alpha_{\eta_2}$. Numerical evaluation of Eq.~(\ref{Eq:dyn1}) for $p_{1,2}(t)$ (see Fig.~\ref{Fig:Dyn_inter}(a)) illustrates that in both cases the lifetimes of {\it geometrically protected} carriers are by the orders of magnitude larger than those in single-group WSM (Fig.~\ref{Fig:Dyn_intra}), the lifetimes of carriers of $W_2$ [inset in Fig.~\ref{Fig:Dyn_inter}(a)], or when the intrinsic node group is unprotected~\cite{SupportingInfo}[$v_1<v_2$, Fig.~\ref{Fig:Dyn_inter}(b)].

{\it Conclusion.} We have shown that peculiar ultra-relativistic band structure of WSM leads to strong suppression of AR due to phase space restrictions imposed by energy and momentum conservation. The only allowed Auger process involves scattering between Weyl nodes with dissimilar carrier dispersion laws %$E_i({\bf k})\neq E_j({\bf k})$
. In comparison to graphene, the carrier lifetimes in realistic WSM are elongated by several orders of magnitude thanks to (1) ''geometrical'' restrictions in phase space and (2) strong screening of Coulomb interaction proportional to the  number of nodes.

The most striking feature of the inter-group AR in WSM is the {\it geometrical protection} of the population inversion in the ''fastest'' node group $W_\star$, where the rate of this process vanishes. The persistent recombination channels due to velocity anisotropy and tilt can be strongly suppressed if $W_\star$ is intrinsic and there are enough carriers outside of it to provide strong screening. As a result, the lifetimes up to $10^{-8}\,\rm{s}$ can be achieved in WSM with {\it geometrically protected} node group. Experimental studies of TaAs support the desired band structure, so the proposed scenario is highly realistic.

%We have shown that peculiar ultra-relativistic band structure of WSM results in strong suppression of AR. The suppression results from phase space restrictions imposed by energy and momentum conservation laws. A relatively weak Auger scattering is allowed between different Weyl nodes due to the dissimilarity of their carrier dispersions, i.e. different Weyl velocity tensors and different tilt. Apart from ''geometrical'' restrictions, extra elongation of carrier lifetimes in WSM appears due to large number of nodes $\eta$ which translates into large dielectric constant.

Our preliminary estimates show that account of dynamic screening further suppresses AR, as the real part of the screening function has a logarithmic singularity at the edge of the excitation cone~\cite{lv2013dielectric}.
%, ${\rm Re}\epsilon(q,\omega) \propto \ln|\omega - q v_0|$
This would result in an extra large log factor in carrier lifetime, order of $\ln ^2 \mathcal G^{-1}$. All these results allow us to consider WSM as promising candidates for long-lasting non-equilibrium states and efficient coherent terahertz emission.

%\begin{acknowledgment}
This work was supported by Grant No. 16-19-10557 of the Russian Science Foundation.
%\end{acknowledgment}

\bibliography{WeylAuger}

\begin{thebibliography}{41}%
\makeatletter
\providecommand \@ifxundefined [1]{%
 \@ifx{#1\undefined}
}%
\providecommand \@ifnum [1]{%
 \ifnum #1\expandafter \@firstoftwo
 \else \expandafter \@secondoftwo
 \fi
}%
\providecommand \@ifx [1]{%
 \ifx #1\expandafter \@firstoftwo
 \else \expandafter \@secondoftwo
 \fi
}%
\providecommand \natexlab [1]{#1}%
\providecommand \enquote  [1]{``#1''}%
\providecommand \bibnamefont  [1]{#1}%
\providecommand \bibfnamefont [1]{#1}%
\providecommand \citenamefont [1]{#1}%
\providecommand \href@noop [0]{\@secondoftwo}%
\providecommand \href [0]{\begingroup \@sanitize@url \@href}%
\providecommand \@href[1]{\@@startlink{#1}\@@href}%
\providecommand \@@href[1]{\endgroup#1\@@endlink}%
\providecommand \@sanitize@url [0]{\catcode `\\12\catcode `\$12\catcode
  `\&12\catcode `\#12\catcode `\^12\catcode `\_12\catcode `\%12\relax}%
\providecommand \@@startlink[1]{}%
\providecommand \@@endlink[0]{}%
\providecommand \url  [0]{\begingroup\@sanitize@url \@url }%
\providecommand \@url [1]{\endgroup\@href {#1}{\urlprefix }}%
\providecommand \urlprefix  [0]{URL }%
\providecommand \Eprint [0]{\href }%
\providecommand \doibase [0]{http://dx.doi.org/}%
\providecommand \selectlanguage [0]{\@gobble}%
\providecommand \bibinfo  [0]{\@secondoftwo}%
\providecommand \bibfield  [0]{\@secondoftwo}%
\providecommand \translation [1]{[#1]}%
\providecommand \BibitemOpen [0]{}%
\providecommand \bibitemStop [0]{}%
\providecommand \bibitemNoStop [0]{.\EOS\space}%
\providecommand \EOS [0]{\spacefactor3000\relax}%
\providecommand \BibitemShut  [1]{\csname bibitem#1\endcsname}%
\let\auto@bib@innerbib\@empty
%</preamble>
\bibitem [{\citenamefont {{A}rmitage}\ \emph {et~al.}(2018)\citenamefont
  {{A}rmitage}, \citenamefont {Mele},\ and\ \citenamefont
  {Vishwanath}}]{Armitage2018}%
  \BibitemOpen
  \bibfield  {author} {\bibinfo {author} {\bibfnamefont {N.~P.}\ \bibnamefont
  {{A}rmitage}}, \bibinfo {author} {\bibfnamefont {E.~J.}\ \bibnamefont
  {Mele}}, \ and\ \bibinfo {author} {\bibfnamefont {A.}~\bibnamefont
  {Vishwanath}},\ }\href {\doibase 10.1103/RevModPhys.90.015001} {\bibfield
  {journal} {\bibinfo  {journal} {Rev. Mod. Phys.}\ }\textbf {\bibinfo {volume}
  {90}},\ \bibinfo {pages} {015001} (\bibinfo {year} {2018})}\BibitemShut
  {NoStop}%
\bibitem [{\citenamefont {Volovik}(2003)}]{volovik2003}%
  \BibitemOpen
  \bibfield  {author} {\bibinfo {author} {\bibfnamefont {G.~E.}\ \bibnamefont
  {Volovik}},\ }\href@noop {} {\emph {\bibinfo {title} {{T}he universe in a
  helium droplet}}},\ Vol.\ \bibinfo {volume} {117}\ (\bibinfo  {publisher}
  {Oxford University Press},\ \bibinfo {year} {2003})\BibitemShut {NoStop}%
\bibitem [{\citenamefont {Elliott}\ and\ \citenamefont
  {Franz}(2015)}]{Elliott_Majorana}%
  \BibitemOpen
  \bibfield  {author} {\bibinfo {author} {\bibfnamefont {S.~R.}\ \bibnamefont
  {Elliott}}\ and\ \bibinfo {author} {\bibfnamefont {M.}~\bibnamefont
  {Franz}},\ }\href {\doibase 10.1103/RevModPhys.87.137} {\bibfield  {journal}
  {\bibinfo  {journal} {Rev. Mod. Phys.}\ }\textbf {\bibinfo {volume} {87}},\
  \bibinfo {pages} {137} (\bibinfo {year} {2015})}\BibitemShut {NoStop}%
\bibitem [{\citenamefont {Novoselov}\ \emph {et~al.}(2005)\citenamefont
  {Novoselov}, \citenamefont {Geim}, \citenamefont {Morozov}, \citenamefont
  {Jiang}, \citenamefont {Katsnelson}, \citenamefont {Grigorieva},
  \citenamefont {{D}ubonos},\ and\ \citenamefont {Firsov}}]{Novoselov2005}%
  \BibitemOpen
  \bibfield  {author} {\bibinfo {author} {\bibfnamefont {K.~S.}\ \bibnamefont
  {Novoselov}}, \bibinfo {author} {\bibfnamefont {A.~K.}\ \bibnamefont {Geim}},
  \bibinfo {author} {\bibfnamefont {S.}~\bibnamefont {Morozov}}, \bibinfo
  {author} {\bibfnamefont {D.}~\bibnamefont {Jiang}}, \bibinfo {author}
  {\bibfnamefont {M.}~\bibnamefont {Katsnelson}}, \bibinfo {author}
  {\bibfnamefont {I.}~\bibnamefont {Grigorieva}}, \bibinfo {author}
  {\bibfnamefont {S.}~\bibnamefont {{D}ubonos}}, \ and\ \bibinfo {author}
  {\bibfnamefont {A.}~\bibnamefont {Firsov}},\ }\href
  {https://www.nature.com/articles/nature04233} {\bibfield  {journal} {\bibinfo
   {journal} {Nature}\ }\textbf {\bibinfo {volume} {438}},\ \bibinfo {pages}
  {197} (\bibinfo {year} {2005})}\BibitemShut {NoStop}%
\bibitem [{\citenamefont {Lv}\ \emph {et~al.}(2015)\citenamefont {Lv},
  \citenamefont {{W}eng}, \citenamefont {Fu}, \citenamefont {{W}ang},
  \citenamefont {Miao}, \citenamefont {Ma}, \citenamefont {Richard},
  \citenamefont {Huang}, \citenamefont {Zhao}, \citenamefont {Chen},
  \citenamefont {Fang}, \citenamefont {{D}ai}, \citenamefont {Qian},\ and\
  \citenamefont {{D}ing}}]{Lv2015TaAs}%
  \BibitemOpen
  \bibfield  {author} {\bibinfo {author} {\bibfnamefont {B.~Q.}\ \bibnamefont
  {Lv}}, \bibinfo {author} {\bibfnamefont {H.~M.}\ \bibnamefont {{W}eng}},
  \bibinfo {author} {\bibfnamefont {B.~B.}\ \bibnamefont {Fu}}, \bibinfo
  {author} {\bibfnamefont {X.~P.}\ \bibnamefont {{W}ang}}, \bibinfo {author}
  {\bibfnamefont {H.}~\bibnamefont {Miao}}, \bibinfo {author} {\bibfnamefont
  {J.}~\bibnamefont {Ma}}, \bibinfo {author} {\bibfnamefont {P.}~\bibnamefont
  {Richard}}, \bibinfo {author} {\bibfnamefont {X.~C.}\ \bibnamefont {Huang}},
  \bibinfo {author} {\bibfnamefont {L.~X.}\ \bibnamefont {Zhao}}, \bibinfo
  {author} {\bibfnamefont {G.~F.}\ \bibnamefont {Chen}}, \bibinfo {author}
  {\bibfnamefont {Z.}~\bibnamefont {Fang}}, \bibinfo {author} {\bibfnamefont
  {X.}~\bibnamefont {{D}ai}}, \bibinfo {author} {\bibfnamefont
  {T.}~\bibnamefont {Qian}}, \ and\ \bibinfo {author} {\bibfnamefont
  {H.}~\bibnamefont {{D}ing}},\ }\href {\doibase 10.1103/PhysRevX.5.031013}
  {\bibfield  {journal} {\bibinfo  {journal} {Phys. Rev. X}\ }\textbf {\bibinfo
  {volume} {5}},\ \bibinfo {pages} {031013} (\bibinfo {year}
  {2015})}\BibitemShut {NoStop}%
\bibitem [{\citenamefont {Xu}\ \emph {et~al.}(2015{\natexlab{a}})\citenamefont
  {Xu}, \citenamefont {Belopolski}, \citenamefont {{A}lidoust}, \citenamefont
  {Neupane}, \citenamefont {Bian}, \citenamefont {Zhang}, \citenamefont
  {Sankar}, \citenamefont {Chang}, \citenamefont {Yuan}, \citenamefont {Lee},
  \citenamefont {Huang}, \citenamefont {Zheng}, \citenamefont {Ma},
  \citenamefont {Sanchez}, \citenamefont {{W}ang}, \citenamefont {Bansil},
  \citenamefont {Chou}, \citenamefont {Shibayev}, \citenamefont {Lin},
  \citenamefont {Jia},\ and\ \citenamefont {Hasan}}]{Xu2015TaAs}%
  \BibitemOpen
  \bibfield  {author} {\bibinfo {author} {\bibfnamefont {S.-Y.}\ \bibnamefont
  {Xu}}, \bibinfo {author} {\bibfnamefont {I.}~\bibnamefont {Belopolski}},
  \bibinfo {author} {\bibfnamefont {N.}~\bibnamefont {{A}lidoust}}, \bibinfo
  {author} {\bibfnamefont {M.}~\bibnamefont {Neupane}}, \bibinfo {author}
  {\bibfnamefont {G.}~\bibnamefont {Bian}}, \bibinfo {author} {\bibfnamefont
  {C.}~\bibnamefont {Zhang}}, \bibinfo {author} {\bibfnamefont
  {R.}~\bibnamefont {Sankar}}, \bibinfo {author} {\bibfnamefont
  {G.}~\bibnamefont {Chang}}, \bibinfo {author} {\bibfnamefont
  {Z.}~\bibnamefont {Yuan}}, \bibinfo {author} {\bibfnamefont {C.-C.}\
  \bibnamefont {Lee}}, \bibinfo {author} {\bibfnamefont {S.-M.}\ \bibnamefont
  {Huang}}, \bibinfo {author} {\bibfnamefont {H.}~\bibnamefont {Zheng}},
  \bibinfo {author} {\bibfnamefont {J.}~\bibnamefont {Ma}}, \bibinfo {author}
  {\bibfnamefont {D.~S.}\ \bibnamefont {Sanchez}}, \bibinfo {author}
  {\bibfnamefont {B.}~\bibnamefont {{W}ang}}, \bibinfo {author} {\bibfnamefont
  {A.}~\bibnamefont {Bansil}}, \bibinfo {author} {\bibfnamefont
  {F.}~\bibnamefont {Chou}}, \bibinfo {author} {\bibfnamefont {P.~P.}\
  \bibnamefont {Shibayev}}, \bibinfo {author} {\bibfnamefont {H.}~\bibnamefont
  {Lin}}, \bibinfo {author} {\bibfnamefont {S.}~\bibnamefont {Jia}}, \ and\
  \bibinfo {author} {\bibfnamefont {M.~Z.}\ \bibnamefont {Hasan}},\ }\href
  {\doibase 10.1126/science.aaa9297} {\bibfield  {journal} {\bibinfo  {journal}
  {Science}\ }\textbf {\bibinfo {volume} {349}},\ \bibinfo {pages} {613}
  (\bibinfo {year} {2015}{\natexlab{a}})}\BibitemShut {NoStop}%
\bibitem [{\citenamefont {Xu}\ \emph {et~al.}(2015{\natexlab{b}})\citenamefont
  {Xu}, \citenamefont {Belopolski}, \citenamefont {Sanchez}, \citenamefont
  {Zhang}, \citenamefont {Chang}, \citenamefont {Guo}, \citenamefont {Bian},
  \citenamefont {Yuan}, \citenamefont {Lu}, \citenamefont {Chang},
  \citenamefont {Shibayev}, \citenamefont {Prokopovych}, \citenamefont
  {{A}lidoust}, \citenamefont {Zheng}, \citenamefont {Lee}, \citenamefont
  {Huang}, \citenamefont {Sankar}, \citenamefont {Chou}, \citenamefont {Hsu},
  \citenamefont {Jeng}, \citenamefont {Bansil}, \citenamefont {Neupert},
  \citenamefont {Strocov}, \citenamefont {Lin}, \citenamefont {Jia},\ and\
  \citenamefont {Hasan}}]{Xu2015TaP}%
  \BibitemOpen
  \bibfield  {author} {\bibinfo {author} {\bibfnamefont {S.-Y.}\ \bibnamefont
  {Xu}}, \bibinfo {author} {\bibfnamefont {I.}~\bibnamefont {Belopolski}},
  \bibinfo {author} {\bibfnamefont {D.~S.}\ \bibnamefont {Sanchez}}, \bibinfo
  {author} {\bibfnamefont {C.}~\bibnamefont {Zhang}}, \bibinfo {author}
  {\bibfnamefont {G.}~\bibnamefont {Chang}}, \bibinfo {author} {\bibfnamefont
  {C.}~\bibnamefont {Guo}}, \bibinfo {author} {\bibfnamefont {G.}~\bibnamefont
  {Bian}}, \bibinfo {author} {\bibfnamefont {Z.}~\bibnamefont {Yuan}}, \bibinfo
  {author} {\bibfnamefont {H.}~\bibnamefont {Lu}}, \bibinfo {author}
  {\bibfnamefont {T.-R.}\ \bibnamefont {Chang}}, \bibinfo {author}
  {\bibfnamefont {P.~P.}\ \bibnamefont {Shibayev}}, \bibinfo {author}
  {\bibfnamefont {M.~L.}\ \bibnamefont {Prokopovych}}, \bibinfo {author}
  {\bibfnamefont {N.}~\bibnamefont {{A}lidoust}}, \bibinfo {author}
  {\bibfnamefont {H.}~\bibnamefont {Zheng}}, \bibinfo {author} {\bibfnamefont
  {C.-C.}\ \bibnamefont {Lee}}, \bibinfo {author} {\bibfnamefont {S.-M.}\
  \bibnamefont {Huang}}, \bibinfo {author} {\bibfnamefont {R.}~\bibnamefont
  {Sankar}}, \bibinfo {author} {\bibfnamefont {F.}~\bibnamefont {Chou}},
  \bibinfo {author} {\bibfnamefont {C.-H.}\ \bibnamefont {Hsu}}, \bibinfo
  {author} {\bibfnamefont {H.-T.}\ \bibnamefont {Jeng}}, \bibinfo {author}
  {\bibfnamefont {A.}~\bibnamefont {Bansil}}, \bibinfo {author} {\bibfnamefont
  {T.}~\bibnamefont {Neupert}}, \bibinfo {author} {\bibfnamefont {V.~N.}\
  \bibnamefont {Strocov}}, \bibinfo {author} {\bibfnamefont {H.}~\bibnamefont
  {Lin}}, \bibinfo {author} {\bibfnamefont {S.}~\bibnamefont {Jia}}, \ and\
  \bibinfo {author} {\bibfnamefont {M.~Z.}\ \bibnamefont {Hasan}},\ }\href
  {\doibase 10.1126/sciadv.1501092} {\bibfield  {journal} {\bibinfo  {journal}
  {Sci. {A}dv.}\ }\textbf {\bibinfo {volume} {1}},\ \bibinfo {pages} {1501092}
  (\bibinfo {year} {2015}{\natexlab{b}})}\BibitemShut {NoStop}%
\bibitem [{\citenamefont {Xu}\ \emph {et~al.}(2015{\natexlab{c}})\citenamefont
  {Xu}, \citenamefont {{A}lidoust}, \citenamefont {Belopolski}, \citenamefont
  {Yuan}, \citenamefont {Bian}, \citenamefont {Chang}, \citenamefont {Zheng},
  \citenamefont {Strocov}, \citenamefont {Sanchez}, \citenamefont {Chang},
  \citenamefont {Zhang}, \citenamefont {Mou}, \citenamefont {{W}u},
  \citenamefont {Huang}, \citenamefont {Lee}, \citenamefont {Huang},
  \citenamefont {{W}ang}, \citenamefont {Bansil}, \citenamefont {Jeng},
  \citenamefont {Neupert}, \citenamefont {Kaminski}, \citenamefont {Lin},
  \citenamefont {Jia},\ and\ \citenamefont {Zahid~Hasan}}]{Xu2015NbAs}%
  \BibitemOpen
  \bibfield  {author} {\bibinfo {author} {\bibfnamefont {S.-Y.}\ \bibnamefont
  {Xu}}, \bibinfo {author} {\bibfnamefont {N.}~\bibnamefont {{A}lidoust}},
  \bibinfo {author} {\bibfnamefont {I.}~\bibnamefont {Belopolski}}, \bibinfo
  {author} {\bibfnamefont {Z.}~\bibnamefont {Yuan}}, \bibinfo {author}
  {\bibfnamefont {G.}~\bibnamefont {Bian}}, \bibinfo {author} {\bibfnamefont
  {T.-R.}\ \bibnamefont {Chang}}, \bibinfo {author} {\bibfnamefont
  {H.}~\bibnamefont {Zheng}}, \bibinfo {author} {\bibfnamefont {V.~N.}\
  \bibnamefont {Strocov}}, \bibinfo {author} {\bibfnamefont {D.~S.}\
  \bibnamefont {Sanchez}}, \bibinfo {author} {\bibfnamefont {G.}~\bibnamefont
  {Chang}}, \bibinfo {author} {\bibfnamefont {C.}~\bibnamefont {Zhang}},
  \bibinfo {author} {\bibfnamefont {D.}~\bibnamefont {Mou}}, \bibinfo {author}
  {\bibfnamefont {Y.}~\bibnamefont {{W}u}}, \bibinfo {author} {\bibfnamefont
  {L.}~\bibnamefont {Huang}}, \bibinfo {author} {\bibfnamefont {C.-C.}\
  \bibnamefont {Lee}}, \bibinfo {author} {\bibfnamefont {S.-M.}\ \bibnamefont
  {Huang}}, \bibinfo {author} {\bibfnamefont {B.}~\bibnamefont {{W}ang}},
  \bibinfo {author} {\bibfnamefont {A.}~\bibnamefont {Bansil}}, \bibinfo
  {author} {\bibfnamefont {H.-T.}\ \bibnamefont {Jeng}}, \bibinfo {author}
  {\bibfnamefont {T.}~\bibnamefont {Neupert}}, \bibinfo {author} {\bibfnamefont
  {A.}~\bibnamefont {Kaminski}}, \bibinfo {author} {\bibfnamefont
  {H.}~\bibnamefont {Lin}}, \bibinfo {author} {\bibfnamefont {S.}~\bibnamefont
  {Jia}}, \ and\ \bibinfo {author} {\bibfnamefont {M.}~\bibnamefont
  {Zahid~Hasan}},\ }\href {http://dx.doi.org/10.1038/nphys3437} {\bibfield
  {journal} {\bibinfo  {journal} {Nat. Phys.}\ }\textbf {\bibinfo {volume}
  {11}},\ \bibinfo {pages} {748} (\bibinfo {year}
  {2015}{\natexlab{c}})}\BibitemShut {NoStop}%
\bibitem [{\citenamefont {Devizorova}\ and\ \citenamefont
  {Volkov}(2017)}]{Devizorova_Weyl}%
  \BibitemOpen
  \bibfield  {author} {\bibinfo {author} {\bibfnamefont {Z.~A.}\ \bibnamefont
  {Devizorova}}\ and\ \bibinfo {author} {\bibfnamefont {V.~A.}\ \bibnamefont
  {Volkov}},\ }\href {\doibase 10.1103/PhysRevB.95.081302} {\bibfield
  {journal} {\bibinfo  {journal} {Phys. Rev. B}\ }\textbf {\bibinfo {volume}
  {95}},\ \bibinfo {pages} {081302} (\bibinfo {year} {2017})}\BibitemShut
  {NoStop}%
\bibitem [{\citenamefont {Liu}\ \emph {et~al.}(2014)\citenamefont {Liu},
  \citenamefont {Zhou}, \citenamefont {Zhang}, \citenamefont {{W}ang},
  \citenamefont {{W}eng}, \citenamefont {Prabhakaran}, \citenamefont {Mo},
  \citenamefont {Shen}, \citenamefont {Fang}, \citenamefont {{D}ai},
  \citenamefont {Hussain},\ and\ \citenamefont {Chen}}]{Liu_Dirac_Na3Bi}%
  \BibitemOpen
  \bibfield  {author} {\bibinfo {author} {\bibfnamefont {Z.~K.}\ \bibnamefont
  {Liu}}, \bibinfo {author} {\bibfnamefont {B.}~\bibnamefont {Zhou}}, \bibinfo
  {author} {\bibfnamefont {Y.}~\bibnamefont {Zhang}}, \bibinfo {author}
  {\bibfnamefont {Z.~J.}\ \bibnamefont {{W}ang}}, \bibinfo {author}
  {\bibfnamefont {H.~M.}\ \bibnamefont {{W}eng}}, \bibinfo {author}
  {\bibfnamefont {D.}~\bibnamefont {Prabhakaran}}, \bibinfo {author}
  {\bibfnamefont {S.-K.}\ \bibnamefont {Mo}}, \bibinfo {author} {\bibfnamefont
  {Z.~X.}\ \bibnamefont {Shen}}, \bibinfo {author} {\bibfnamefont
  {Z.}~\bibnamefont {Fang}}, \bibinfo {author} {\bibfnamefont {X.}~\bibnamefont
  {{D}ai}}, \bibinfo {author} {\bibfnamefont {Z.}~\bibnamefont {Hussain}}, \
  and\ \bibinfo {author} {\bibfnamefont {Y.~L.}\ \bibnamefont {Chen}},\ }\href
  {\doibase 10.1126/science.1245085} {\bibfield  {journal} {\bibinfo  {journal}
  {Science}\ }\textbf {\bibinfo {volume} {343}},\ \bibinfo {pages} {864}
  (\bibinfo {year} {2014})}\BibitemShut {NoStop}%
\bibitem [{\citenamefont {Stander}\ \emph {et~al.}(2009)\citenamefont
  {Stander}, \citenamefont {Huard},\ and\ \citenamefont
  {Goldhaber-Gordon}}]{Stander_KleinGraphene}%
  \BibitemOpen
  \bibfield  {author} {\bibinfo {author} {\bibfnamefont {N.}~\bibnamefont
  {Stander}}, \bibinfo {author} {\bibfnamefont {B.}~\bibnamefont {Huard}}, \
  and\ \bibinfo {author} {\bibfnamefont {D.}~\bibnamefont {Goldhaber-Gordon}},\
  }\href {\doibase 10.1103/PhysRevLett.102.026807} {\bibfield  {journal}
  {\bibinfo  {journal} {Phys. Rev. Lett.}\ }\textbf {\bibinfo {volume} {102}},\
  \bibinfo {pages} {026807} (\bibinfo {year} {2009})}\BibitemShut {NoStop}%
\bibitem [{\citenamefont {{W}ang}\ \emph {et~al.}(2013)\citenamefont {{W}ang},
  \citenamefont {{W}ong}, \citenamefont {Shytov}, \citenamefont {Brar},
  \citenamefont {Choi}, \citenamefont {{W}u}, \citenamefont {{T}sai},
  \citenamefont {Regan}, \citenamefont {Zettl}, \citenamefont {Kawakami},
  \citenamefont {Louie}, \citenamefont {Levitov},\ and\ \citenamefont
  {Crommie}}]{Levitov_AtomicCollapse}%
  \BibitemOpen
  \bibfield  {author} {\bibinfo {author} {\bibfnamefont {Y.}~\bibnamefont
  {{W}ang}}, \bibinfo {author} {\bibfnamefont {D.}~\bibnamefont {{W}ong}},
  \bibinfo {author} {\bibfnamefont {A.~V.}\ \bibnamefont {Shytov}}, \bibinfo
  {author} {\bibfnamefont {V.~W.}\ \bibnamefont {Brar}}, \bibinfo {author}
  {\bibfnamefont {S.}~\bibnamefont {Choi}}, \bibinfo {author} {\bibfnamefont
  {Q.}~\bibnamefont {{W}u}}, \bibinfo {author} {\bibfnamefont {H.-Z.}\
  \bibnamefont {{T}sai}}, \bibinfo {author} {\bibfnamefont {W.}~\bibnamefont
  {Regan}}, \bibinfo {author} {\bibfnamefont {A.}~\bibnamefont {Zettl}},
  \bibinfo {author} {\bibfnamefont {R.~K.}\ \bibnamefont {Kawakami}}, \bibinfo
  {author} {\bibfnamefont {S.~G.}\ \bibnamefont {Louie}}, \bibinfo {author}
  {\bibfnamefont {L.~S.}\ \bibnamefont {Levitov}}, \ and\ \bibinfo {author}
  {\bibfnamefont {M.~F.}\ \bibnamefont {Crommie}},\ }\href {\doibase
  10.1126/science.1234320} {\bibfield  {journal} {\bibinfo  {journal}
  {Science}\ }\textbf {\bibinfo {volume} {340}},\ \bibinfo {pages} {734}
  (\bibinfo {year} {2013})}\BibitemShut {NoStop}%
\bibitem [{\citenamefont {Zhang}\ \emph {et~al.}(2016)\citenamefont {Zhang},
  \citenamefont {Xu}, \citenamefont {Belopolski}, \citenamefont {Yuan},
  \citenamefont {Lin}, \citenamefont {{T}ong}, \citenamefont {Bian},
  \citenamefont {{A}lidoust}, \citenamefont {Lee}, \citenamefont {Huang} \emph
  {et~al.}}]{zhang_ABJAnomaly}%
  \BibitemOpen
  \bibfield  {author} {\bibinfo {author} {\bibfnamefont {C.-L.}\ \bibnamefont
  {Zhang}}, \bibinfo {author} {\bibfnamefont {S.-Y.}\ \bibnamefont {Xu}},
  \bibinfo {author} {\bibfnamefont {I.}~\bibnamefont {Belopolski}}, \bibinfo
  {author} {\bibfnamefont {Z.}~\bibnamefont {Yuan}}, \bibinfo {author}
  {\bibfnamefont {Z.}~\bibnamefont {Lin}}, \bibinfo {author} {\bibfnamefont
  {B.}~\bibnamefont {{T}ong}}, \bibinfo {author} {\bibfnamefont
  {G.}~\bibnamefont {Bian}}, \bibinfo {author} {\bibfnamefont {N.}~\bibnamefont
  {{A}lidoust}}, \bibinfo {author} {\bibfnamefont {C.-C.}\ \bibnamefont {Lee}},
  \bibinfo {author} {\bibfnamefont {S.-M.}\ \bibnamefont {Huang}},  \emph
  {et~al.},\ }\href {https://www.nature.com/articles/ncomms10735} {\bibfield
  {journal} {\bibinfo  {journal} {Nature Commun.}\ }\textbf {\bibinfo {volume}
  {7}},\ \bibinfo {pages} {10735} (\bibinfo {year} {2016})}\BibitemShut
  {NoStop}%
\bibitem [{\citenamefont {Gooth}\ \emph {et~al.}(2017)\citenamefont {Gooth},
  \citenamefont {Niemann}, \citenamefont {Meng}, \citenamefont {Grushin},
  \citenamefont {Landsteiner}, \citenamefont {Gotsmann}, \citenamefont
  {Menges}, \citenamefont {Schmidt}, \citenamefont {Shekhar}, \citenamefont
  {Suss}, \citenamefont {Huhne}, \citenamefont {Rellinghaus}, \citenamefont
  {Felser}, \citenamefont {Yan},\ and\ \citenamefont {Nielsch}}]{Gooth2017}%
  \BibitemOpen
  \bibfield  {author} {\bibinfo {author} {\bibfnamefont {J.}~\bibnamefont
  {Gooth}}, \bibinfo {author} {\bibfnamefont {A.~C.}\ \bibnamefont {Niemann}},
  \bibinfo {author} {\bibfnamefont {T.}~\bibnamefont {Meng}}, \bibinfo {author}
  {\bibfnamefont {A.~G.}\ \bibnamefont {Grushin}}, \bibinfo {author}
  {\bibfnamefont {K.}~\bibnamefont {Landsteiner}}, \bibinfo {author}
  {\bibfnamefont {B.}~\bibnamefont {Gotsmann}}, \bibinfo {author}
  {\bibfnamefont {F.}~\bibnamefont {Menges}}, \bibinfo {author} {\bibfnamefont
  {M.}~\bibnamefont {Schmidt}}, \bibinfo {author} {\bibfnamefont
  {C.}~\bibnamefont {Shekhar}}, \bibinfo {author} {\bibfnamefont
  {V.}~\bibnamefont {Suss}}, \bibinfo {author} {\bibfnamefont {R.}~\bibnamefont
  {Huhne}}, \bibinfo {author} {\bibfnamefont {B.}~\bibnamefont {Rellinghaus}},
  \bibinfo {author} {\bibfnamefont {C.}~\bibnamefont {Felser}}, \bibinfo
  {author} {\bibfnamefont {B.}~\bibnamefont {Yan}}, \ and\ \bibinfo {author}
  {\bibfnamefont {K.}~\bibnamefont {Nielsch}},\ }\href {\doibase
  10.1038/nature23005} {\bibfield  {journal} {\bibinfo  {journal} {Nature}\
  }\textbf {\bibinfo {volume} {547}},\ \bibinfo {pages} {324} (\bibinfo {year}
  {2017})}\BibitemShut {NoStop}%
\bibitem [{\citenamefont {Liang}\ \emph {et~al.}(2015)\citenamefont {Liang},
  \citenamefont {Gibson}, \citenamefont {{A}li}, \citenamefont {Liu},
  \citenamefont {Cava},\ and\ \citenamefont
  {Ong}}]{BackscatteringSuppressionWeyl}%
  \BibitemOpen
  \bibfield  {author} {\bibinfo {author} {\bibfnamefont {T.}~\bibnamefont
  {Liang}}, \bibinfo {author} {\bibfnamefont {Q.}~\bibnamefont {Gibson}},
  \bibinfo {author} {\bibfnamefont {M.~N.}\ \bibnamefont {{A}li}}, \bibinfo
  {author} {\bibfnamefont {M.}~\bibnamefont {Liu}}, \bibinfo {author}
  {\bibfnamefont {R.}~\bibnamefont {Cava}}, \ and\ \bibinfo {author}
  {\bibfnamefont {N.}~\bibnamefont {Ong}},\ }\href
  {https://www.nature.com/articles/nmat4143} {\bibfield  {journal} {\bibinfo
  {journal} {Nat. Mater.}\ }\textbf {\bibinfo {volume} {14}},\ \bibinfo {pages}
  {280} (\bibinfo {year} {2015})}\BibitemShut {NoStop}%
\bibitem [{\citenamefont {Mayorov}\ \emph {et~al.}(2011)\citenamefont
  {Mayorov}, \citenamefont {Gorbachev}, \citenamefont {Morozov}, \citenamefont
  {Britnell}, \citenamefont {Jalil}, \citenamefont {Ponomarenko}, \citenamefont
  {Blake}, \citenamefont {Novoselov}, \citenamefont {{W}atanabe}, \citenamefont
  {{T}aniguchi},\ and\ \citenamefont
  {Geim}}]{BakcscatteringSuppressionGraphene}%
  \BibitemOpen
  \bibfield  {author} {\bibinfo {author} {\bibfnamefont {A.~S.}\ \bibnamefont
  {Mayorov}}, \bibinfo {author} {\bibfnamefont {R.~V.}\ \bibnamefont
  {Gorbachev}}, \bibinfo {author} {\bibfnamefont {S.~V.}\ \bibnamefont
  {Morozov}}, \bibinfo {author} {\bibfnamefont {L.}~\bibnamefont {Britnell}},
  \bibinfo {author} {\bibfnamefont {R.}~\bibnamefont {Jalil}}, \bibinfo
  {author} {\bibfnamefont {L.~A.}\ \bibnamefont {Ponomarenko}}, \bibinfo
  {author} {\bibfnamefont {P.}~\bibnamefont {Blake}}, \bibinfo {author}
  {\bibfnamefont {K.~S.}\ \bibnamefont {Novoselov}}, \bibinfo {author}
  {\bibfnamefont {K.}~\bibnamefont {{W}atanabe}}, \bibinfo {author}
  {\bibfnamefont {T.}~\bibnamefont {{T}aniguchi}}, \ and\ \bibinfo {author}
  {\bibfnamefont {A.~K.}\ \bibnamefont {Geim}},\ }\href {\doibase
  10.1021/nl200758b} {\bibfield  {journal} {\bibinfo  {journal} {Nano Lett.}\
  }\textbf {\bibinfo {volume} {11}},\ \bibinfo {pages} {2396} (\bibinfo {year}
  {2011})}\BibitemShut {NoStop}%
\bibitem [{\citenamefont {Kibis}(2014)}]{BackscatteringKibis}%
  \BibitemOpen
  \bibfield  {author} {\bibinfo {author} {\bibfnamefont {O.~V.}\ \bibnamefont
  {Kibis}},\ }\href {http://stacks.iop.org/0295-5075/107/i=5/a=57003}
  {\bibfield  {journal} {\bibinfo  {journal} {Europhys. Lett.}\ }\textbf
  {\bibinfo {volume} {107}},\ \bibinfo {pages} {57003} (\bibinfo {year}
  {2014})}\BibitemShut {NoStop}%
\bibitem [{\citenamefont {Keldysh}(1986)}]{keldysh1986EHL}%
  \BibitemOpen
  \bibfield  {author} {\bibinfo {author} {\bibfnamefont {L.}~\bibnamefont
  {Keldysh}},\ }\href
  {https://www.tandfonline.com/doi/pdf/10.1080/00107518608211022} {\bibfield
  {journal} {\bibinfo  {journal} {Contemp. Phys.}\ }\textbf {\bibinfo {volume}
  {27}},\ \bibinfo {pages} {395} (\bibinfo {year} {1986})}\BibitemShut
  {NoStop}%
\bibitem [{\citenamefont {{T}riola}\ \emph {et~al.}(2017)\citenamefont
  {{T}riola}, \citenamefont {Pertsova}, \citenamefont {Markiewicz},\ and\
  \citenamefont {Balatsky}}]{Triola_Insulator}%
  \BibitemOpen
  \bibfield  {author} {\bibinfo {author} {\bibfnamefont {C.}~\bibnamefont
  {{T}riola}}, \bibinfo {author} {\bibfnamefont {A.}~\bibnamefont {Pertsova}},
  \bibinfo {author} {\bibfnamefont {R.~S.}\ \bibnamefont {Markiewicz}}, \ and\
  \bibinfo {author} {\bibfnamefont {A.~V.}\ \bibnamefont {Balatsky}},\ }\href
  {\doibase 10.1103/PhysRevB.95.205410} {\bibfield  {journal} {\bibinfo
  {journal} {Phys. Rev. B}\ }\textbf {\bibinfo {volume} {95}},\ \bibinfo
  {pages} {205410} (\bibinfo {year} {2017})}\BibitemShut {NoStop}%
\bibitem [{\citenamefont {Lindner}\ \emph {et~al.}(2011)\citenamefont
  {Lindner}, \citenamefont {Refael},\ and\ \citenamefont
  {Galitski}}]{Floquet_Insulator}%
  \BibitemOpen
  \bibfield  {author} {\bibinfo {author} {\bibfnamefont {N.~H.}\ \bibnamefont
  {Lindner}}, \bibinfo {author} {\bibfnamefont {G.}~\bibnamefont {Refael}}, \
  and\ \bibinfo {author} {\bibfnamefont {V.}~\bibnamefont {Galitski}},\ }\href
  {https://www.nature.com/articles/nphys1926} {\bibfield  {journal} {\bibinfo
  {journal} {Nat. Phys.}\ }\textbf {\bibinfo {volume} {7}},\ \bibinfo {pages}
  {490} (\bibinfo {year} {2011})}\BibitemShut {NoStop}%
\bibitem [{\citenamefont {Iveland}\ \emph {et~al.}(2013)\citenamefont
  {Iveland}, \citenamefont {Martinelli}, \citenamefont {Peretti}, \citenamefont
  {Speck},\ and\ \citenamefont {{W}eisbuch}}]{Iveland_AugerDroop}%
  \BibitemOpen
  \bibfield  {author} {\bibinfo {author} {\bibfnamefont {J.}~\bibnamefont
  {Iveland}}, \bibinfo {author} {\bibfnamefont {L.}~\bibnamefont {Martinelli}},
  \bibinfo {author} {\bibfnamefont {J.}~\bibnamefont {Peretti}}, \bibinfo
  {author} {\bibfnamefont {J.~S.}\ \bibnamefont {Speck}}, \ and\ \bibinfo
  {author} {\bibfnamefont {C.}~\bibnamefont {{W}eisbuch}},\ }\href {\doibase
  10.1103/PhysRevLett.110.177406} {\bibfield  {journal} {\bibinfo  {journal}
  {Phys. Rev. Lett.}\ }\textbf {\bibinfo {volume} {110}},\ \bibinfo {pages}
  {177406} (\bibinfo {year} {2013})}\BibitemShut {NoStop}%
\bibitem [{\citenamefont {Morozov}\ \emph {et~al.}(2017)\citenamefont
  {Morozov}, \citenamefont {Rumyantsev}, \citenamefont {Fadeev}, \citenamefont
  {Zholudev}, \citenamefont {Kudryavtsev}, \citenamefont {{A}ntonov},
  \citenamefont {Kadykov}, \citenamefont {{D}ubinov}, \citenamefont
  {Mikhailov}, \citenamefont {{D}voretsky},\ and\ \citenamefont
  {Gavrilenko}}]{Morozov_StimulatedHgTe}%
  \BibitemOpen
  \bibfield  {author} {\bibinfo {author} {\bibfnamefont {S.~V.}\ \bibnamefont
  {Morozov}}, \bibinfo {author} {\bibfnamefont {V.~V.}\ \bibnamefont
  {Rumyantsev}}, \bibinfo {author} {\bibfnamefont {M.~A.}\ \bibnamefont
  {Fadeev}}, \bibinfo {author} {\bibfnamefont {M.~S.}\ \bibnamefont
  {Zholudev}}, \bibinfo {author} {\bibfnamefont {K.~E.}\ \bibnamefont
  {Kudryavtsev}}, \bibinfo {author} {\bibfnamefont {A.~V.}\ \bibnamefont
  {{A}ntonov}}, \bibinfo {author} {\bibfnamefont {A.~M.}\ \bibnamefont
  {Kadykov}}, \bibinfo {author} {\bibfnamefont {A.~A.}\ \bibnamefont
  {{D}ubinov}}, \bibinfo {author} {\bibfnamefont {N.~N.}\ \bibnamefont
  {Mikhailov}}, \bibinfo {author} {\bibfnamefont {S.~A.}\ \bibnamefont
  {{D}voretsky}}, \ and\ \bibinfo {author} {\bibfnamefont {V.~I.}\ \bibnamefont
  {Gavrilenko}},\ }\href {\doibase 10.1063/1.4996966} {\bibfield  {journal}
  {\bibinfo  {journal} {{A}ppl. Phys. Lett.}\ }\textbf {\bibinfo {volume}
  {111}},\ \bibinfo {pages} {192101} (\bibinfo {year} {2017})}\BibitemShut
  {NoStop}%
\bibitem [{\citenamefont {Adams}(1986)}]{Adams_strain}%
  \BibitemOpen
  \bibfield  {author} {\bibinfo {author} {\bibfnamefont {A.~R.}\ \bibnamefont
  {Adams}},\ }\href {\doibase 10.1049/el:19860171} {\bibfield  {journal}
  {\bibinfo  {journal} {Electronics Letters}\ }\textbf {\bibinfo {volume}
  {22}},\ \bibinfo {pages} {249} (\bibinfo {year} {1986})}\BibitemShut
  {NoStop}%
\bibitem [{\citenamefont {Cragg}\ and\ \citenamefont
  {Efros}(2010)}]{Efros_Suppression}%
  \BibitemOpen
  \bibfield  {author} {\bibinfo {author} {\bibfnamefont {G.~E.}\ \bibnamefont
  {Cragg}}\ and\ \bibinfo {author} {\bibfnamefont {A.~L.}\ \bibnamefont
  {Efros}},\ }\href {\doibase 10.1021/nl903592h} {\bibfield  {journal}
  {\bibinfo  {journal} {Nano Lett.}\ }\textbf {\bibinfo {volume} {10}},\
  \bibinfo {pages} {313} (\bibinfo {year} {2010})}\BibitemShut {NoStop}%
\bibitem [{\citenamefont {Nawrocki}\ \emph {et~al.}(1995)\citenamefont
  {Nawrocki}, \citenamefont {Rubo}, \citenamefont {Lascaray},\ and\
  \citenamefont {Coquillat}}]{Magnetic_suppression}%
  \BibitemOpen
  \bibfield  {author} {\bibinfo {author} {\bibfnamefont {M.}~\bibnamefont
  {Nawrocki}}, \bibinfo {author} {\bibfnamefont {Y.~G.}\ \bibnamefont {Rubo}},
  \bibinfo {author} {\bibfnamefont {J.~P.}\ \bibnamefont {Lascaray}}, \ and\
  \bibinfo {author} {\bibfnamefont {D.}~\bibnamefont {Coquillat}},\ }\href
  {\doibase 10.1103/PhysRevB.52.R2241} {\bibfield  {journal} {\bibinfo
  {journal} {Phys. Rev. B}\ }\textbf {\bibinfo {volume} {52}},\ \bibinfo
  {pages} {R2241} (\bibinfo {year} {1995})}\BibitemShut {NoStop}%
\bibitem [{\citenamefont {{A}rnold}\ \emph {et~al.}(2016)\citenamefont
  {{A}rnold}, \citenamefont {Shekhar}, \citenamefont {{W}u}, \citenamefont
  {Sun}, \citenamefont {dos Reis}, \citenamefont {Kumar}, \citenamefont
  {Naumann}, \citenamefont {{A}jeesh}, \citenamefont {Schmidt}, \citenamefont
  {Grushin}, \citenamefont {Bardarson}, \citenamefont {Baenitz}, \citenamefont
  {Sokolov}, \citenamefont {Borrmann}, \citenamefont {Nicklas}, \citenamefont
  {Felser}, \citenamefont {Hassinger},\ and\ \citenamefont {Yan}}]{Arnold2016}%
  \BibitemOpen
  \bibfield  {author} {\bibinfo {author} {\bibfnamefont {F.}~\bibnamefont
  {{A}rnold}}, \bibinfo {author} {\bibfnamefont {C.}~\bibnamefont {Shekhar}},
  \bibinfo {author} {\bibfnamefont {S.-C.}\ \bibnamefont {{W}u}}, \bibinfo
  {author} {\bibfnamefont {Y.}~\bibnamefont {Sun}}, \bibinfo {author}
  {\bibfnamefont {R.~D.}\ \bibnamefont {dos Reis}}, \bibinfo {author}
  {\bibfnamefont {N.}~\bibnamefont {Kumar}}, \bibinfo {author} {\bibfnamefont
  {M.}~\bibnamefont {Naumann}}, \bibinfo {author} {\bibfnamefont {M.~O.}\
  \bibnamefont {{A}jeesh}}, \bibinfo {author} {\bibfnamefont {M.}~\bibnamefont
  {Schmidt}}, \bibinfo {author} {\bibfnamefont {A.~G.}\ \bibnamefont
  {Grushin}}, \bibinfo {author} {\bibfnamefont {J.~H.}\ \bibnamefont
  {Bardarson}}, \bibinfo {author} {\bibfnamefont {M.}~\bibnamefont {Baenitz}},
  \bibinfo {author} {\bibfnamefont {D.}~\bibnamefont {Sokolov}}, \bibinfo
  {author} {\bibfnamefont {H.}~\bibnamefont {Borrmann}}, \bibinfo {author}
  {\bibfnamefont {M.}~\bibnamefont {Nicklas}}, \bibinfo {author} {\bibfnamefont
  {C.}~\bibnamefont {Felser}}, \bibinfo {author} {\bibfnamefont
  {E.}~\bibnamefont {Hassinger}}, \ and\ \bibinfo {author} {\bibfnamefont
  {B.}~\bibnamefont {Yan}},\ }\href {http://dx.doi.org/10.1038/ncomms11615}
  {\bibfield  {journal} {\bibinfo  {journal} {Nat. Commun.}\ }\textbf {\bibinfo
  {volume} {7}},\ \bibinfo {pages} {11615} (\bibinfo {year}
  {2016})}\BibitemShut {NoStop}%
\bibitem [{\citenamefont {Hu}\ \emph {et~al.}(2016)\citenamefont {Hu},
  \citenamefont {Liu}, \citenamefont {Graf}, \citenamefont {Radmanesh},
  \citenamefont {{A}dams}, \citenamefont {Chuang}, \citenamefont {{W}ang},
  \citenamefont {Chiorescu}, \citenamefont {{W}ei}, \citenamefont {Spinu},\
  and\ \citenamefont {Mao}}]{Hu2016}%
  \BibitemOpen
  \bibfield  {author} {\bibinfo {author} {\bibfnamefont {J.}~\bibnamefont
  {Hu}}, \bibinfo {author} {\bibfnamefont {J.~Y.}\ \bibnamefont {Liu}},
  \bibinfo {author} {\bibfnamefont {D.}~\bibnamefont {Graf}}, \bibinfo {author}
  {\bibfnamefont {S.~M.~A.}\ \bibnamefont {Radmanesh}}, \bibinfo {author}
  {\bibfnamefont {D.~J.}\ \bibnamefont {{A}dams}}, \bibinfo {author}
  {\bibfnamefont {A.}~\bibnamefont {Chuang}}, \bibinfo {author} {\bibfnamefont
  {Y.}~\bibnamefont {{W}ang}}, \bibinfo {author} {\bibfnamefont
  {I.}~\bibnamefont {Chiorescu}}, \bibinfo {author} {\bibfnamefont
  {J.}~\bibnamefont {{W}ei}}, \bibinfo {author} {\bibfnamefont
  {L.}~\bibnamefont {Spinu}}, \ and\ \bibinfo {author} {\bibfnamefont {Z.~Q.}\
  \bibnamefont {Mao}},\ }\href {http://dx.doi.org/10.1038/srep18674} {\bibfield
   {journal} {\bibinfo  {journal} {Sci. Rep.}\ }\textbf {\bibinfo {volume}
  {6}},\ \bibinfo {pages} {18674} (\bibinfo {year} {2016})}\BibitemShut
  {NoStop}%
\bibitem [{\citenamefont {Klotz}\ \emph {et~al.}(2016)\citenamefont {Klotz},
  \citenamefont {{W}u}, \citenamefont {Shekhar}, \citenamefont {Sun},
  \citenamefont {Schmidt}, \citenamefont {Nicklas}, \citenamefont {Baenitz},
  \citenamefont {Uhlarz}, \citenamefont {{W}osnitza}, \citenamefont {Felser},\
  and\ \citenamefont {Yan}}]{Klotz2016}%
  \BibitemOpen
  \bibfield  {author} {\bibinfo {author} {\bibfnamefont {J.}~\bibnamefont
  {Klotz}}, \bibinfo {author} {\bibfnamefont {S.-C.}\ \bibnamefont {{W}u}},
  \bibinfo {author} {\bibfnamefont {C.}~\bibnamefont {Shekhar}}, \bibinfo
  {author} {\bibfnamefont {Y.}~\bibnamefont {Sun}}, \bibinfo {author}
  {\bibfnamefont {M.}~\bibnamefont {Schmidt}}, \bibinfo {author} {\bibfnamefont
  {M.}~\bibnamefont {Nicklas}}, \bibinfo {author} {\bibfnamefont
  {M.}~\bibnamefont {Baenitz}}, \bibinfo {author} {\bibfnamefont
  {M.}~\bibnamefont {Uhlarz}}, \bibinfo {author} {\bibfnamefont
  {J.}~\bibnamefont {{W}osnitza}}, \bibinfo {author} {\bibfnamefont
  {C.}~\bibnamefont {Felser}}, \ and\ \bibinfo {author} {\bibfnamefont
  {B.}~\bibnamefont {Yan}},\ }\href {\doibase 10.1103/PhysRevB.93.121105}
  {\bibfield  {journal} {\bibinfo  {journal} {Phys. Rev. B}\ }\textbf {\bibinfo
  {volume} {93}},\ \bibinfo {pages} {121105} (\bibinfo {year}
  {2016})}\BibitemShut {NoStop}%
\bibitem [{\citenamefont {Emtage}(1976)}]{Emtage_AR_Lead_tin}%
  \BibitemOpen
  \bibfield  {author} {\bibinfo {author} {\bibfnamefont {P.~R.}\ \bibnamefont
  {Emtage}},\ }\href {\doibase 10.1063/1.322975} {\bibfield  {journal}
  {\bibinfo  {journal} {J. {A}ppl. Phys.}\ }\textbf {\bibinfo {volume} {47}},\
  \bibinfo {pages} {2565} (\bibinfo {year} {1976})}\BibitemShut {NoStop}%
\bibitem [{\citenamefont {Ziep}\ and\ \citenamefont {Mocker}(1980)}]{Ziep1980}%
  \BibitemOpen
  \bibfield  {author} {\bibinfo {author} {\bibfnamefont {O.}~\bibnamefont
  {Ziep}}\ and\ \bibinfo {author} {\bibfnamefont {M.}~\bibnamefont {Mocker}},\
  }\href {\doibase 10.1002/pssb.2220980113} {\bibfield  {journal} {\bibinfo
  {journal} {Phys. Status Solidi B}\ }\textbf {\bibinfo {volume} {98}},\
  \bibinfo {pages} {133} (\bibinfo {year} {1980})}\BibitemShut {NoStop}%
\bibitem [{\citenamefont {{D}irac}(1930)}]{Dirac1930}%
  \BibitemOpen
  \bibfield  {author} {\bibinfo {author} {\bibfnamefont {P.~A.~M.}\
  \bibnamefont {{D}irac}},\ }\href {\doibase 10.1017/S0305004100016091}
  {\bibfield  {journal} {\bibinfo  {journal} {Mathematical Proceedings of the
  Cambridge Philosophical Society}\ }\textbf {\bibinfo {volume} {26}},\
  \bibinfo {pages} {361–375} (\bibinfo {year} {1930})}\BibitemShut {NoStop}%
\bibitem [{\citenamefont {Fritz}\ \emph {et~al.}(2008)\citenamefont {Fritz},
  \citenamefont {Schmalian}, \citenamefont {Muller},\ and\ \citenamefont
  {Sachdev}}]{Sachdev_QuantumCritical}%
  \BibitemOpen
  \bibfield  {author} {\bibinfo {author} {\bibfnamefont {L.}~\bibnamefont
  {Fritz}}, \bibinfo {author} {\bibfnamefont {J.}~\bibnamefont {Schmalian}},
  \bibinfo {author} {\bibfnamefont {M.}~\bibnamefont {Muller}}, \ and\ \bibinfo
  {author} {\bibfnamefont {S.}~\bibnamefont {Sachdev}},\ }\href {\doibase
  10.1103/PhysRevB.78.085416} {\bibfield  {journal} {\bibinfo  {journal} {Phys.
  Rev. B}\ }\textbf {\bibinfo {volume} {78}},\ \bibinfo {pages} {085416}
  (\bibinfo {year} {2008})}\BibitemShut {NoStop}%
\bibitem [{\citenamefont {Rana}(2007)}]{Rana_Auger_2007}%
  \BibitemOpen
  \bibfield  {author} {\bibinfo {author} {\bibfnamefont {F.}~\bibnamefont
  {Rana}},\ }\href {\doibase 10.1103/PhysRevB.76.155431} {\bibfield  {journal}
  {\bibinfo  {journal} {Phys. Rev. B}\ }\textbf {\bibinfo {volume} {76}},\
  \bibinfo {pages} {155431} (\bibinfo {year} {2007})}\BibitemShut {NoStop}%
\bibitem [{\citenamefont {{W}inzer}\ and\ \citenamefont
  {Malic}(2012)}]{Malic-Auger}%
  \BibitemOpen
  \bibfield  {author} {\bibinfo {author} {\bibfnamefont {T.}~\bibnamefont
  {{W}inzer}}\ and\ \bibinfo {author} {\bibfnamefont {E.}~\bibnamefont
  {Malic}},\ }\href {https://arxiv.org/pdf/1204.5650.pdf} {\bibfield  {journal}
  {\bibinfo  {journal} {Phys. Rev. B}\ }\textbf {\bibinfo {volume} {85}},\
  \bibinfo {pages} {241404} (\bibinfo {year} {2012})}\BibitemShut {NoStop}%
\bibitem [{\citenamefont {{T}omadin}\ \emph {et~al.}(2013)\citenamefont
  {{T}omadin}, \citenamefont {Brida}, \citenamefont {Cerullo}, \citenamefont
  {Ferrari},\ and\ \citenamefont {Polini}}]{Tomadin-theory}%
  \BibitemOpen
  \bibfield  {author} {\bibinfo {author} {\bibfnamefont {A.}~\bibnamefont
  {{T}omadin}}, \bibinfo {author} {\bibfnamefont {D.}~\bibnamefont {Brida}},
  \bibinfo {author} {\bibfnamefont {G.}~\bibnamefont {Cerullo}}, \bibinfo
  {author} {\bibfnamefont {A.~C.}\ \bibnamefont {Ferrari}}, \ and\ \bibinfo
  {author} {\bibfnamefont {M.}~\bibnamefont {Polini}},\ }\href {\doibase
  10.1103/PhysRevB.88.035430} {\bibfield  {journal} {\bibinfo  {journal} {Phys.
  Rev. B}\ }\textbf {\bibinfo {volume} {88}},\ \bibinfo {pages} {035430}
  (\bibinfo {year} {2013})}\BibitemShut {NoStop}%
\bibitem [{\citenamefont {{A}lymov}\ \emph {et~al.}(2018)\citenamefont
  {{A}lymov}, \citenamefont {Vyurkov}, \citenamefont {Ryzhii}, \citenamefont
  {Satou},\ and\ \citenamefont {Svintsov}}]{Svintsov2018}%
  \BibitemOpen
  \bibfield  {author} {\bibinfo {author} {\bibfnamefont {G.}~\bibnamefont
  {{A}lymov}}, \bibinfo {author} {\bibfnamefont {V.}~\bibnamefont {Vyurkov}},
  \bibinfo {author} {\bibfnamefont {V.}~\bibnamefont {Ryzhii}}, \bibinfo
  {author} {\bibfnamefont {A.}~\bibnamefont {Satou}}, \ and\ \bibinfo {author}
  {\bibfnamefont {D.}~\bibnamefont {Svintsov}},\ }\href {\doibase
  10.1103/PhysRevB.97.205411} {\bibfield  {journal} {\bibinfo  {journal} {Phys.
  Rev. B}\ }\textbf {\bibinfo {volume} {97}},\ \bibinfo {pages} {205411}
  (\bibinfo {year} {2018})}\BibitemShut {NoStop}%
\bibitem [{Note1()}]{Note1}%
  \BibitemOpen
  \bibinfo {note} {We note that Eq.~(\ref {AR-rate}) incorporates both {A}uger
  recombination and generation processes, therefore it vanishes in
  equilibrium.}\BibitemShut {Stop}%
\bibitem [{Sup()}]{SupportingInfo}%
  \BibitemOpen
  \href@noop {} {\ }\bibinfo {note} {See online supplemental material for I.
  Evaluation of {W}SM polarizability II. Factorization of {A}R rate III.
  Evaluation of $\mathcal G$ for various shapes of dispersion surfaces IV.
  Evaluation of $\mathcal S$ and screening functions in various pumping regimes
  V. {D}iscussion of relaxation in multi-group WSM}\BibitemShut {NoStop}%
\bibitem [{\citenamefont {Huang}\ \emph {et~al.}(2018)\citenamefont {Huang},
  \citenamefont {Sanderson}, \citenamefont {{T}ian}, \citenamefont {Chen},
  \citenamefont {{{W}}ang},\ and\ \citenamefont
  {Zhang}}]{Huang_2018_HotCarrier}%
  \BibitemOpen
  \bibfield  {author} {\bibinfo {author} {\bibfnamefont {S.}~\bibnamefont
  {Huang}}, \bibinfo {author} {\bibfnamefont {M.}~\bibnamefont {Sanderson}},
  \bibinfo {author} {\bibfnamefont {J.}~\bibnamefont {{T}ian}}, \bibinfo
  {author} {\bibfnamefont {Q.}~\bibnamefont {Chen}}, \bibinfo {author}
  {\bibfnamefont {F.}~\bibnamefont {{{W}}ang}}, \ and\ \bibinfo {author}
  {\bibfnamefont {C.}~\bibnamefont {Zhang}},\ }\href
  {http://stacks.iop.org/0022-3727/51/i=1/a=015101} {\bibfield  {journal}
  {\bibinfo  {journal} {J. Phys. {D}: {A}ppl. Phys.}\ }\textbf {\bibinfo
  {volume} {51}},\ \bibinfo {pages} {015101} (\bibinfo {year}
  {2018})}\BibitemShut {NoStop}%
\bibitem [{\citenamefont {{Ruan}}\ \emph {et~al.}(2016)\citenamefont {{Ruan}},
  \citenamefont {{Jian}}, \citenamefont {{Yao}}, \citenamefont {{Zhang}},
  \citenamefont {{Zhang}},\ and\ \citenamefont {{Xing}}}]{Ruan2016HgTe}%
  \BibitemOpen
  \bibfield  {author} {\bibinfo {author} {\bibfnamefont {J.}~\bibnamefont
  {{Ruan}}}, \bibinfo {author} {\bibfnamefont {S.-K.}\ \bibnamefont {{Jian}}},
  \bibinfo {author} {\bibfnamefont {H.}~\bibnamefont {{Yao}}}, \bibinfo
  {author} {\bibfnamefont {H.}~\bibnamefont {{Zhang}}}, \bibinfo {author}
  {\bibfnamefont {S.-C.}\ \bibnamefont {{Zhang}}}, \ and\ \bibinfo {author}
  {\bibfnamefont {D.}~\bibnamefont {{Xing}}},\ }\href {\doibase
  10.1038/ncomms11136} {\bibfield  {journal} {\bibinfo  {journal} {Nat.
  Commun.}\ }\textbf {\bibinfo {volume} {7}},\ \bibinfo {pages} {11136}
  (\bibinfo {year} {2016})}\BibitemShut {NoStop}%
\bibitem [{\citenamefont {Lv}\ and\ \citenamefont
  {Zhang}(2013)}]{lv2013dielectric}%
  \BibitemOpen
  \bibfield  {author} {\bibinfo {author} {\bibfnamefont {M.}~\bibnamefont
  {Lv}}\ and\ \bibinfo {author} {\bibfnamefont {S.-C.}\ \bibnamefont {Zhang}},\
  }\href {https://www.worldscientific.com/doi/abs/10.1142/S0217979213501774}
  {\bibfield  {journal} {\bibinfo  {journal} {International Journal of Modern
  Physics B}\ }\textbf {\bibinfo {volume} {27}},\ \bibinfo {pages} {1350177}
  (\bibinfo {year} {2013})}\BibitemShut {NoStop}%
\end{thebibliography}%

\newpage

\begin{widetext}
\renewcommand{\theequation}{S\arabic{equation}}
\renewcommand{\thefigure}{S\arabic{figure}}
\setcounter{equation}{0}
\setcounter{figure}{0}
\section*{Supplemental material}

%%%%%%%%%%%%%%%%%%%%%%%%%%%%%%%%%%%%%%%%%%%%%%%
\section{Imaginary parts of polarizabilities in Weyl semimetals}
The RPA polarizability due to electron transitions between $s$-th and $s'$-th bands is given by~\cite{lv2013dielectric}
\begin{equation}
\label{P_RPA}
\Pi_{ss'}(\omega,\textbf{q})=\frac{\eta}{V}\sum\limits_{\textbf{k}}\frac{f_s(E_{s{\bf k}})-f_{s'}(E_{s'{\bf k}'})}{\omega+E_{s{\bf k}}-E_{s'{\bf k}'}+i0}\mathcal{I}_{ss'}(\textbf{k},\textbf{k}'),
\end{equation}
where $s,s'=\pm1$, $\eta$ is the number of Weyl points of a given type, $\textbf{k}'=\textbf{k}+\textbf{q}$, $f_s(E_{s{\bf k}})$ is the Fermi function with quasi-Fermi energy $\mu_s$, $\mathcal{I}_{ss'}(\textbf{k},\textbf{k}')$ is the squared overlap integral between states $\{s\textbf{k}\}$ and $\{s'\textbf{k}'\}$. The imaginary part of polarizability is obtained from~(\ref{P_RPA}) with Sokhotski theorem
\begin{equation}
\label{ImP_RPA}
{\rm Im} \Pi_{ss'}(\omega,\textbf{q})=-\eta\pi\int \frac{d\textbf{k}}{{(2\pi)^3}}\mathcal{I}_{ss'}(\textbf{k},\textbf{k}')\left[f_s(E_{s{\bf k}})-f_{s'}(E_{s'{\bf k}'})\right]\delta(\omega+E_{s{\bf k}}-E_{s'{\bf k}'}).
\end{equation}
The spectrum of a prototypical WSM is $E_{s{\bf k}}=s v_0 k$, while the squared overlap integral takes on the form
\begin{equation}
\mathcal{I}_{ss'}(\textbf{k},\textbf{k}')=\frac{1}{2}\left(1+ss'\frac{\textbf{k}\textbf{k}'}{k k'}\right).
\end{equation}
For brevity, we set from now on the Weyl velocity to unity, $v_0\equiv 1$. This allows us to transform~(\ref{ImP_RPA}):
\begin{multline}
{\rm Im} \Pi^{(0)}_{ss'}(\omega,\textbf{q})=-\frac{\eta s s'}{16\pi q}\int\limits_0^{+\infty}dk\int\limits_{|k-q|}^{k+q}dk'\delta(\omega+s k-s' k')\times\\
\times[(sk+s'k')^2-q^2][f_s(s k)-f_{s'}(s' k')].
\end{multline}
The latter form was achieved by passing from integration over the spherical angles to the integration over the modulus of $k'=|\textbf{k}+\textbf{q}|$. Considering separately the regions $k>q$ and $k<q$ and performing the change of variable $k=(\omega + q x)/2$, we obtain the final expressions for inter- and intraband parts of WSM polarizability ($\omega_{\bf q}=v_0q$):
\begin{gather}
\label{ImPss_reduced}
{\rm Im}\Pi^{(0)}_{ss}(\omega,q)=\Theta(\omega_{\bf q}-\omega)F_{ss}(\omega,q)\\
\label{Fss}
F_{ss}(\omega,q)=-s\frac{\omega_{\bf q}^2}{32\pi v_0^3}\int\limits_1^{+\infty}dx (x^2-1)\left[f_s\left(s\frac{\omega_{\bf q} x-\omega}{2}\right)-f_s\left(s\frac{\omega_q x+\omega}{2}\right)\right]\\
\label{ImPvc_reduced}
{\rm Im}\Pi^{(0)}_{-s,s}(\omega,q)=\Theta(s\omega-\omega_{\bf q})F_{-ss}(\omega,q)\\
\label{F-+}
F_{-s,s}(\omega,q)=-\frac{\omega_q^2}{32\pi v_0^3}\int\limits_{-1}^{1}dx (1-x^2)\left[f_{-s}\left(-s\frac{\omega_{\bf q} x+\omega}{2}\right)-f_s\left(s\frac{\omega-\omega_{\bf q} x}{2}\right)\right]
\end{gather}
%%%%%%%%%%%%%%%%%%%%%%%%%%%%%%%%%%%%%%%%%%%%%%%
\section{Auger recombination rate factorization}
The general expression for AR rate between Weyl nodes $i$ and $j$ belonging to the groups $W_n$ and $W_{n'}$  has the form
\begin{equation}
\label{AR-common}
{\mathcal R}^{(s)}_{ij} = 4 \sum\limits_{{\bf q}\omega}  {\rm Im}\Pi_{-+}^{(i)}(\omega,{\bf q})
\frac{|V_0(q)|^2}{|\epsilon(\omega,{\bf q})|^2} {\rm Im}\Pi_{ss}^{(j)}(\omega,{\bf q}) \mathcal{N}_B(\omega),
\end{equation}
where $\mathcal{N}_B(\omega)=n_B(\omega - \Delta\mu_{eh}) - n_B(\omega)$ and $\Pi_{-+}^{(i)}(\omega,\textbf{q})=\Pi_{-1,1}^{(i)}(\omega,\textbf{q})$. Using Eq.~(6) of the main text and expressions~(\ref{ImPss_reduced})-(\ref{F-+}), we can re-write the imaginary parts of polarizabilities $\Pi_{ss'}^{(i,j)}$ for anisotropic Weyl nodes as
\begin{gather}
{\rm Im}\Pi_{-+}^{(i)}(\omega,{\bf q})=\Theta(\omega-\tilde{\omega}_l({\bf q}))F_{-+}^{(0)}(\omega_i({\bf q}),q_i)\\
{\rm Im}\Pi_{ss}^{(j)}(\omega,{\bf q})=\Theta(\tilde{\omega}_j({\bf q})-\omega)F_{ss}^{(0)}(\omega_j({\bf q}),q_j)\\
\label{omega_i}
\omega_{i}({\bf q})=\omega-{\bf u}_{n} g_{n}^{(i)}{\bf q}\\
\tilde{\omega}_i({\bf q})=|\hat{v}_ng_n^{(i)}{\bf q}|+{\bf u}_n g_n^{(i)}{\bf q}\\
\label{q_i}
{\bf q}_i=v_n^{-1}\hat{v}_n g_n^{(i)}{\bf q},
\end{gather}
The velocity of prototypical WSM is taken as average velocity, $v_{n,n'}=\sqrt[3]{|{\rm det}\hat{v}_{n,n'}|}$, while $\omega_j({\bf q}), \tilde{\omega}_j({\bf q}), {\bf q}_j$ are the same as~(\ref{omega_i})-(\ref{q_i}) with replacements $i\rightarrow j, n\rightarrow n'$. We note that deriving these expressions in the presence of tilt, ${\bf u}_n \neq 0$, we can neglect  the tilt in the arguments of distribution functions and, hence, in Eqs.~(\ref{Fss}) and~(\ref{F-+}). This is justified by the relatives smallness of tilt in type-I WSM, $|{\bf u}_n| \ll ||\hat{v}_n||$. As a result, the expression for ARR is transformed into
\begin{equation}
{\mathcal R}^{(s)}_{ij} = 4  \int\Theta(\tilde{\omega}_j({\bf q})-\tilde{\omega_i}({\bf q}))\frac{d{\bf q}}{(2\pi)^3}\int\limits_{\tilde{\omega}_i({\bf q})}^{\tilde{\omega}_j({\bf q})}\frac{d\omega}{2\pi} F_{-+}^{(0)}(\omega_i({\bf q}),q_i)
\frac{|V_0(q)|^2}{|\epsilon(\omega,{\bf q})|^2} F_{ss}^{(0)}(\omega_j({\bf q}),q_j) \mathcal{N}_B(\omega),
\end{equation}
To proceed further, we perform the integration variable change ${\bf q} \rightarrow {\bf q}_i$ which transforms the dispersion term, associated with the Weyl velocity, into isotropic one $|\hat{v}_i{\bf q}|=v_n|{\bf q}_i|$. Henceforth, we rename the new variable as ${\bf q}$, and the expression for the ARR takes the form 
\begin{equation}
\label{AR-shifted}
{\mathcal R}^{(s)}_{ij} = 4 \int\limits_0^{+\infty}\frac{q^2 dq}{(2\pi)^4}\int\frac{\Theta(\Delta({\bf e_q}))d{\bf e_q}}{|v_n[g_n^{(i)}]^{-1}{\hat v}_n^{-1}{\bf e_q}|^4}\int\limits_{\tilde{\omega}_i({\bf Q})}^{\tilde{\omega}_j({\bf Q})} F_{-+}^{(0)}(\omega_i({\bf Q}),q)
\frac{|V_0(q)|^2}{|\epsilon(\omega,{\bf Q})|^2} F_{ss}^{(0)}(\omega_j({\bf Q}),q_{ij}) \mathcal{N}_B(\omega),
\end{equation}
where ${\bf Q}=v_n [g_n^{(i)}]^{-1} \hat{v}_{n}^{-1}{\bf q}$, $q_{ij}=|(v_n/v_{n'})\hat{v}_{n'}g_{ij}\hat{v}_n^{-1}{\bf q}|$, and $g_{ij}=g_{n'}^{(j)}[g_n^{(i)}]^{-1}$. In~(\ref{AR-shifted}) we have factorized ${\bf q}$-integration and ``bare'' Coulomb interaction into radial $q=|{\bf q}|$ and angular ${\bf e_q}={\bf q}/q$ parts 
\begin{equation}
|V_0({\bf Q})|^2=\frac{|V_0(q)|^2}{|v_n[g_n^{(i)}]^{-1}{\hat v}_n^{-1}{\bf e_q}|^4}.
\end{equation}
The domain of the allowed $(\omega,{\bf e_q})$ is determined by 
\begin{gather}
\Delta({\bf e_q}) =\frac{\tilde{\omega}_j({\bf Q})-\tilde{\omega}_i({\bf Q})}{v_n q}= 
\left|{\hat v}_{n'}g_{ij}{\hat v}_{n}^{-1}{\bf e_q}\right|-1-({\bf u}_n-{\bf u}_{n'}g_{ij}){\hat v}_n^{-1}{\bf e_q}\\
\Delta({\bf e_q})>0,
\end{gather}
where $\Delta({\bf e_q})$ is its thickness over $\omega$ in the units of $\omega_{\bf q}=v_n q$, being nonzero due to the dissimilarity in the nodes' $i$ and $j$ dispersion.

We limit ourselves to the case when this difference is small. Then, due to $\Delta({\bf e_q})\ll1$, $\omega_{i,j}({\bf Q})\approx \omega_{\bf q}$ and $v_{n'}\approx v_n=v_0$. Since the dependence of the integrand in~(\ref{AR-shifted}) on $\omega$ is determined by the smooth behaviour of the distribution functions, in~(\ref{AR-shifted}) we can neglect the difference between $\omega_{i,j}({\bf Q})$ and $\omega_q$ everywhere, except the variation range of $\omega$
\begin{equation}
{\mathcal R}^{(s)}_{ij} = 4 \int\limits_0^{+\infty}\frac{q^2 dq}{(2\pi)^4}\int\limits_{\Delta({\bf e_q})>0}\frac{\Delta({\bf e_q})d{\bf e_q}}{|v_n[g_n^{(i)}]^{-1}{\hat v}_n^{-1}{\bf e_q}|^4} F_{-+}^{(0)}(\omega_{\bf q},q)
\frac{|V_0(q)|^2}{|\epsilon(\omega_{\bf q},{\bf Q})|^2} F_{ss}^{(0)}(\omega_{\bf q},q_{ij}) \mathcal{N}_B(\omega).
\end{equation}
We note, that the given expression for ARR is valid only if the dielectric function has no singularities in the frequency range $[\tilde{\omega}_i({\bf Q}),\tilde{\omega}_j({\bf Q})]$. Otherwise, the $\epsilon(\omega)$ dependence should be taken into account when calculating the integral over $\omega$. Analogically, due to the small difference in nodes' $i$ and $j$ dispersions, $q_{ij}\approx q$. Thus, integration over $q$ and ${\bf e_q}$ can be carried separately, so that $R^{(s)}_{ij}$ is factorized into the {\em geometric} and {\em statistical} parts
\begin{gather}
\label{AR-factor}
R^{(s)}_{ij}=\mathcal{G}^{(s)}_{ij} \mathcal{S}_{ij}^{(s)}\\
\mathcal{G}^{(s)}_{ij} = 
\int\limits_{\Delta({\bf e_q})>0}  \frac{\Delta({\bf e_q})d{\bf e_q}}{|v_n[g_n^{(i)}]^{-1}{\hat v}_n^{-1}{\bf e_q}|^4}\\
\mathcal{S}^{(s)}_{ij} = \int \frac{\omega_{\bf q}d^3{\bf q}}{16\pi^5} \frac{|V_0( q)|^2}{|\epsilon(\omega_{\bf q},q)|^2}
F_{-+}^{(i)}(\omega_{\bf q},q) F^{(j)}_{ss}(\omega_{\bf q},q) \mathcal{N}_B(\omega_{\bf q}).
\end{gather}
Strictly speaking, such factorization is valid only when $\epsilon(\omega,{\bf q})=\varkappa$. In general case, $\epsilon(\omega_{\bf q},{\bf Q})$ contains the combinations of $q$ and $|v_n[g_n^{(i)}]^{-1}{\hat v}_n^{-1}{\bf e_q}|^2$, so that factorization becomes possible only with the dielectric function, averaged over ${\bf e_q} $ directions: $\epsilon(\omega_{\bf q},q)=\langle\epsilon(\omega_{\bf q},{\bf Q})\rangle_{\bf e_q}$.

%%%%%%%%%%%%%%%%%%%%%%%%%%%%%%%%%%%%%%%%%%%%%%%
\section{Geometry factor}
The direct evaluation of geometrical factor is most convenient if $\mathcal{G}_{ij}$ is written directly in terms of Weyl parameters of the nodes $i$ and $j$
\begin{gather}
\mathcal{G}_{ij} = 
\int\limits_{\Delta({\bf e_q})>0}  \frac{\Delta({\bf e_q})d{\bf e_q}}{|v_n [{\hat v}^{(i)}]^{-1}{\bf e_q}|^4}, \\
\Delta({\bf e_q}) = 
\left|{\hat v}^{(j)}[{\hat v}^{(i)}]^{-1}{\bf e_q}\right|-1-({\bf u}^{(i)}-{\bf u}^{(j)})[{\hat v}^{(i)}]^{-1}{\bf e_q},
\end{gather}
where $\hat{v}^{(i,j)}=\hat{v}_{n,n'}g^{(i,j)}_{n,n'}$, ${\bf u}^{(i,j)}={\bf u}_{n,n'}g^{(i,j)}_{n,n'}$. In further calculations, we shall use both Cartesian ${\bf e_q}=(e_x,e_y,e_z)$ and spherical coordinates ${\bf e_q}=(\sin{\theta}\cos{\varphi},\sin{\theta}\sin{\varphi},\cos{\theta})$ for the unit vector ${\bf e_q}$.

%%%%%%%%%%%%%%%%%%%%%%%%%%%%

\subsection{Intra-group AR enabled by in-plane velocity anisotropy}

In the $W_1$ node group of TaAs, the Weyl velocity tensors has the form
\begin{equation}
\hat{v}^{(i)}={\rm diag}(v_x\cos{\varphi_i}-v_y\sin{\varphi_i}, v_y\cos{\varphi_i}+v_x\sin{\varphi_i}, v_z),
\end{equation}
where $\varphi_i=0,\pi/2,\pi,3\pi/2$. We evaluate the geometrical factor $\mathcal{G}_{xy}$ using the pair of Weyl points with $\varphi_i=0$ and $\varphi_j=\pi/2$ as an example. The explicit form of their velocity tensors is
\begin{gather}
\hat{v}^{(i)}=\hat{v}^{(xy)}={\rm diag}(v_x,v_y,v_z)\\
\hat{v}^{(j)}=\hat{v}^{(xy_\perp)}={\rm diag}(-v_y,v_x,v_z).
\end{gather}
The ''anisotropy term'' in the denominator of geometrical factor has the form
\begin{equation}
|v_n [{\hat v}^{(xy)}]^{-1}{\bf e_q}|^4=(v_x v_y v_z )^{4/3}\left(\frac{\cos^2\theta}{v_z^2}+\sin^2\theta\left(\frac{\cos^2\varphi}{v_x^2}+\frac{\sin^2\varphi}{v_y^2}\right)\right)^2.
\end{equation}
The domain of allowed $(\omega,{\bf e_q})$ (in the coordinates turning the dispersion of the $i$-th node into isotropic one) represents the intersect of the interior of sphere and exterior of ellipsoid.  The local thickness of this domain
\begin{equation}
\label{AreaAnis}
\Delta({\bf e_q})=\sqrt{e_x^2 \frac{v_y^2}{v_x^2}+e_y^2\frac{v_x^2}{v_y^2}+e_z^2}-1.
\end{equation}
Introducing the eccentricity of constant energy ellipsoid in the $xy$-plane
\begin{equation}
\epsilon_{xy}=\sqrt{1-\frac{v_x^2}{v_y^2}},
\end{equation}
we rewrite (\ref{AreaAnis}) as
\begin{equation}
\Delta({\bf e_q})=\sqrt{1-e_y^2\epsilon_{xy}^2+e_x^2 \frac{\epsilon_{xy}^2}{1-\epsilon_{xy}^2}}-1.
\end{equation}
As mentioned before, we assume the Weyl parameters of the valleys involved to be almost identical. In particular, this implies weak anisotropy in the $xy$ plane: $\epsilon_{xy}\ll 1$, which is equivalent to $|v_x-v_y|\ll v_{\perp}=(v_x+v_y)/2$. The thickness of allowed domain, in this limit, reduces to
\begin{gather}
\Delta({\bf e_q})=\frac{\epsilon_{xy}^2}{2}\sin^2{\theta}\cos{2\varphi}\\
\Delta({\bf e_q})>0 \rightarrow -\frac{\pi}{4}<\varphi<\frac{\pi}{4},
\end{gather}
and will be small as far as $\epsilon_{xy}$ is small. The denominator of geometry factor, in the same limit, takes a simpler form
\begin{equation}
|v_n {\hat v}_{xy}^{-1}{\bf e_q}|^4\approx\frac{(1-\epsilon_{z}^2 \sin^2{\theta})^2}{(1-\epsilon_{z}^2)^{4/3}},
\end{equation}
where $\epsilon_{z}=\sqrt{1-(v_{z}/v_{\perp})^2}$. Then
\begin{equation}
\mathcal{G}_{xy}=(1-\epsilon_z^2)\frac{\epsilon_{xy}^2}{2}\int\limits_{-\frac{\pi}{4}}^{\frac{\pi}{4}} \cos{2\varphi} d\varphi \int\limits_{0}^{\pi} \frac{\sin^3{\theta} d\theta
}{(1-\epsilon_{\parallel}^2 \sin^2{\theta})^2}.
\end{equation}
Direct evaluation of the latter integral leads us to the final result
\begin{gather}
\mathcal{G}_{xy} = \frac{|v_x-v_y|}{v_{\perp}} g(1-v_z^2/v_\perp^2)\left(\frac{v_z}{v_{\perp}}\right)^{5/3}\\
\mathcal{G}_{xy}=(1-\epsilon_z^2)^{5/6} g(\epsilon_{z}^2)\frac{\epsilon_{xy}^2}{2}\\
g(x)=\frac{\sqrt{x(1-x)}+(2x-1)\arctan{\sqrt{\frac{x}{1-x}}}}{(1-x)x^{3/2}}.
\end{gather}

%%%%%%%%%%%%%%%%%
\subsection{Inter-group AR}
The recombination between different node groups is enabled already by the difference of absolute values of Weyl velocities. For this reason, we omit tilt and consider velocity tensor as isotropic
\begin{gather}
\hat{v}^{(i)}=\hat{v}^{(1)}=v_1 \delta_{lm}\\
\hat{v}^{(j)}=\hat{v}^{(2)}=v_2 \delta_{lm},
\end{gather}
and the process is allowed only if $v_2>v_1$. In this recombination mechanism, no symmetry relations between nodes are required and the smallness of the domain of allowed $(\omega,{\bf q})$ is guaranteed by proximity of $v_2$ and $v_1$. Therefore
\begin{gather}
|v_n [{\hat v}^{(1)}]^{-1}{\bf e_q}|^4=1\\
\Delta({\bf e_q})=2\frac{v_2-v_1}{v_1+v_2}.
\end{gather}
Then
\begin{equation}
\mathcal{G}_{\circledcirc}=\frac{8\pi(v_2-v_1)\theta(v_2-v_1)}{v_1+v_2}.
\end{equation}

%%%%%%%%%%%%%%%%%
\subsection{Tilt-enabled AR}
To highlight the effect of tilt-enabled AR, we omit the velocity anisotropy from consideration and take the velocity tensors of $i$-th and $j$-th nodes to be
\begin{equation}
\hat{v}^{(i,j)}=v_0\delta_{lm}.
\end{equation}
The tilt vectors of the nodes can be, however, different
\begin{gather}
\textbf{u}^{(i)}=(u^{(i)}_{x},u^{(i)}_{y},u^{(i)}_{z})\\
\textbf{u}^{(j)}=(u^{(j)}_{x},u^{(j)}_{y},u^{(j)}_{z}).
\end{gather}
The indices $i$ and $j$ can belong to different groups ($W_n$ and $W_{n'}$) or to one group. In the latter case, the tilt vectors are related via symmetry operations. In particular, for a pair of nodes from Sec. IIIA: 
\begin{gather}
\textbf{u}^{(i)}=(u_x^{(i)},u_y^{(i)},u_z^{(i)})\\
\textbf{u}^{(j)}=(-u_y^{(i)},u_x^{(i)},u_z^{(i)}).
\end{gather}
We note that in the case of tilt-enabled AR, the vectors $\textbf{u}_{(j)}$ and $\textbf{u}^{(i)}$ should not necessarily be close to each other. Indeed, the smallness of domain of allowed frequencies $\omega$ is governed by
\begin{equation}
\Delta({\bf e_q})=\frac{(\textbf{u}^{(j)}-\textbf{u}^{(i)}){\bf e_q}}{v_0}
\end{equation}
and is enabled by the smallness of $\textbf{u}^{(i,j)}$ as compared with Weyl velocity $v_0$. Using the Weyl parameters from Sec. IIIA, we obtain
\begin{equation}
|v_n [{\hat v}^{(i)}]^{-1}{\bf e_q}|^4=1.
\end{equation}
The expression for geometrical part takes on the form
\begin{equation}
\mathcal{G}_t = 
\int\limits_{\textbf{u}^{(j)}{\bf e_q}>\textbf{u}^{(i)}{\bf e_q}} \frac{(\textbf{u}^{(j)}-\textbf{u}^{(i)}){\bf e_q}}{v_0} d{\bf e_q}.
\end{equation}
To evaluate the latter, we direct the $z$-axis along $\textbf{u}^{(j)}-\textbf{u}^{(i)}$, then $(\textbf{u}^{(j)}-\textbf{u}^{(i)}){\bf e_q}=|\textbf{u}^{(j)}-\textbf{u}^{(i)}|\cos\theta$. The resulting expression reads
\begin{equation}
\mathcal{G}_t=\frac{\pi|\textbf{u}^{(j)}-\textbf{u}^{(i)}|}{v_0}.
\end{equation}

%%%%%%%%%%%%%%%%%%%%%%%%%%%%%%%%%%%%%%%%%%%%%
\section{Statistical factor and screening}

In this section, we give explicit expressions for statistical factors corresponding to various initial occupations and pumping regimes and provide the particular form of distributions and dielectric functions used to calculate them.

%%%%%%%%%%%%%%%%%%%%%%%%%%%%%%%
\subsection{Intra-group recombination. Sub-exponential relaxation in the intrinsic node group.}
Symmetric pumping of intrinsic node leads to equal quasi-Fermi levels of electrons and holes which we denote as $\delta\mu$, $\mu_c = \delta\mu$, $\mu_v = - \delta\mu$. In the degenerate limit $\mu_c,|\mu_v|\gg T$, the distribution functions can be regarded as step-wise:
\begin{gather}
f_c(E)=\Theta(\delta\mu-E),\\
f_v(E)=\Theta(-\delta\mu-E).
\end{gather}
The non-equilibrium carrier densities are related to quasi-Fermi level via
\begin{equation}
n=p=\frac{\eta}{6\pi^2}\frac{\delta\mu^3}{v_0^3}.
\end{equation}
The screening function in the degenerate limit reads
\begin{equation}
\epsilon(0,q)=1+\frac{4\alpha_{\eta}}{\pi}\frac{\delta\mu^2}{\omega_q^2},
\end{equation}
where $\alpha_{\eta}=\frac{\eta e^2}{v_0\varkappa}$ is the effective fine structure constant which determines the strength of screening. In WSM, $\alpha_{\eta}\gg1$ due to large number of nodes.

With the above assumptions, the statistical factor can be evaluated analytically:
\begin{equation}
\mathcal{S}=C_{1d}(\alpha_{\eta})v_0 \frac{p^{4/3}}{\eta^{4/3}}
\end{equation}
\begin{multline}
C_{1d}(\alpha)=\frac{\alpha^2 6^{1/3}}{4\pi^{10/3}}\left[8\alpha(\alpha+\pi)\ln{\alpha}-(3\pi^2+12\pi\alpha+8\alpha^2)\ln{\left(\alpha+\frac{\pi}{4}\right)}\right.\\
+\left.\pi(3\pi+4\alpha)\ln{(\pi+\alpha)}+2\sqrt{\alpha\pi^3}{\rm arctan}\left(\frac{\sqrt{\pi\alpha}}{\pi+2\alpha}\right)-\frac{9\pi^2}{4}+2\pi\alpha\right],
\end{multline}
where the screening function $C_{1d}(\alpha)$, presented in Fig.~\ref{Fig:CsubI}, has the limiting forms
\begin{gather}
C_{1d}(\alpha)\propto\frac{3(8\ln{2}-3)6^{1/3}\alpha^2}{16\pi^{4/3}}~(\alpha\rightarrow 0),\\
C_{1d}(\alpha)\rightarrow \frac{499}{7680}\pi^{2/3} 6^{1/3}~(\alpha\rightarrow \infty).
\end{gather}

\begin{figure}[h]
	\includegraphics[width=150mm]{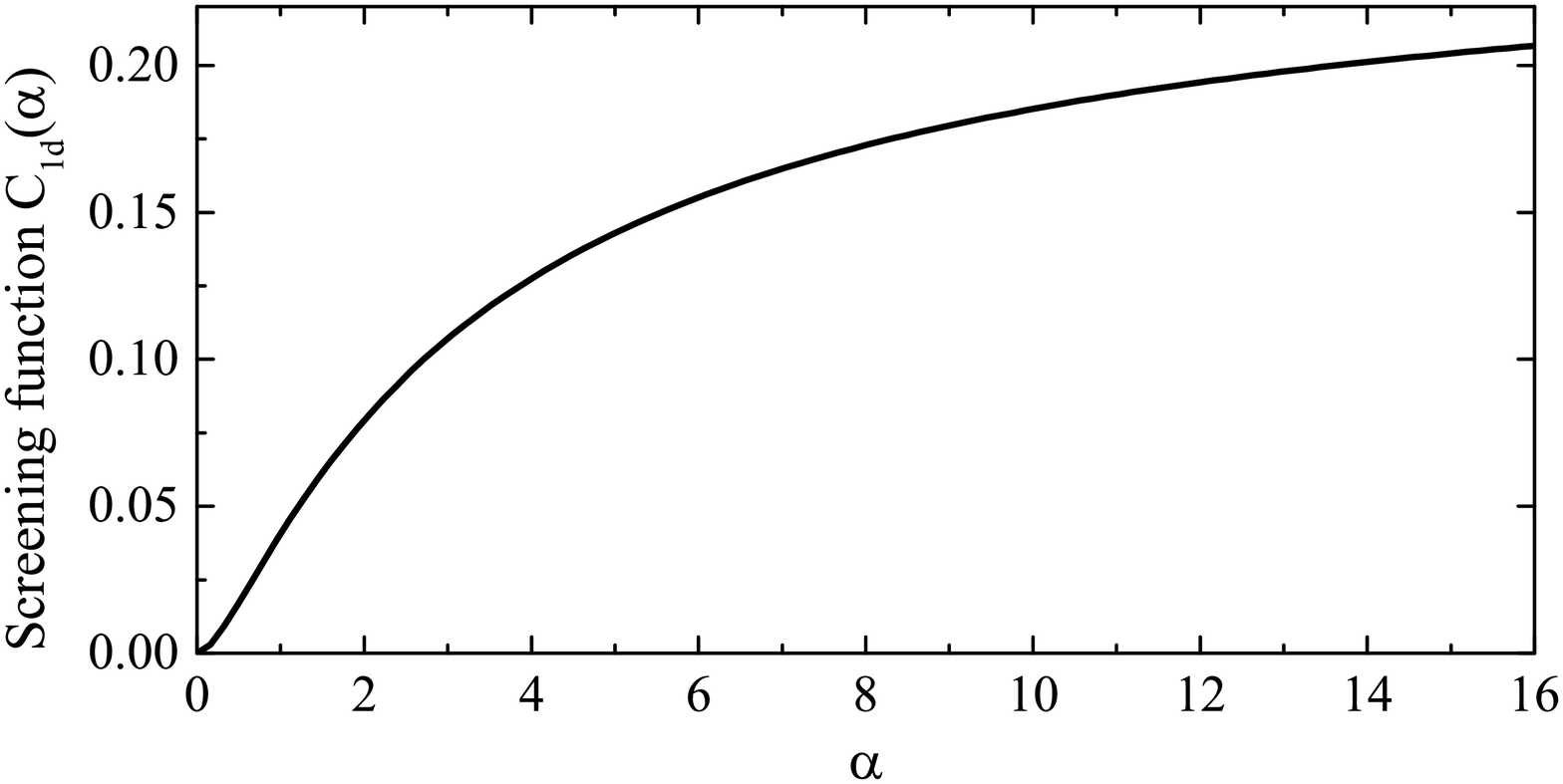}
	\caption{\label{Fig:CsubI}
		Screening function for sub-exponential relaxation {\em via} intra-group processes in the intrinsic nodes.}
\end{figure}

%%%%%%%%%%%%%%%%%%%%%%%%%%%%%
\subsection{Intra-group recombination. Exponential relaxation in the intrinsic node group.}
When quasi-Fermi level of pumped carriers is low $\delta\mu\ll T$, the carrier distribution is non-degenerate and the non-equilibrium density becomes a linear function of $\delta\mu$:

%Distribution functions:
%\begin{gather}
%f_c(E)\approx \frac{1}{\exp\left(\frac{E}{T}\right)+1}+\frac{\delta\mu}{T} \frac{1}{4{\rm ch}^2\left(\frac{E}{2T}\right)}\\
%f_v(E)\approx \frac{1}{\exp\left(-\frac{E}{T}\right)+1}+\frac{\delta\mu}{T}\frac{1}{4{\rm ch}^2\left(\frac{E}{2T}\right)}
%\end{gather}
\begin{equation}
n=p=\frac{\eta}{12}\frac{T^2\delta\mu}{v_0^3}.
\end{equation}
The dielectric constant is governed both by thermal and non-equilibrium carriers:
\begin{equation}
\epsilon(0,q)=1+\frac{2\pi\alpha_{\eta}}{3}\frac{T^2}{\omega_q^2}\left[1+\frac{12\ln{2}}{\pi^2}\frac{\delta\mu}{T}\right]
\end{equation}
The statistical factor is, apparently, linear in excess carrier density:
\begin{equation}
\mathcal{S}=C_{1n}(\alpha_{\eta})T \frac{p}{\eta}.
\end{equation}
The screening function in this limit, depicted in Fig.~\ref{Fig:CexpI}, is not expressed via elementary functions
\begin{equation}
C_{1n}(\alpha)=\frac{3\alpha^2}{256\pi^4}\int\limits_0^{+\infty}\frac{y^7 dy}{(y^2+2\pi\alpha/3)^2}\left[\int\limits_1^{+\infty}\frac{(x^2-1)dx}{{\rm ch}(y/2)+{\rm ch}(xy/2)}\right]\left[\int\limits_{-1}^{1}\frac{(1-x^2)dx}{{\rm ch}(y/2)+{\rm ch}(xy/2)}\right]
\end{equation}

\begin{figure}[h]
	\includegraphics[width=150mm]{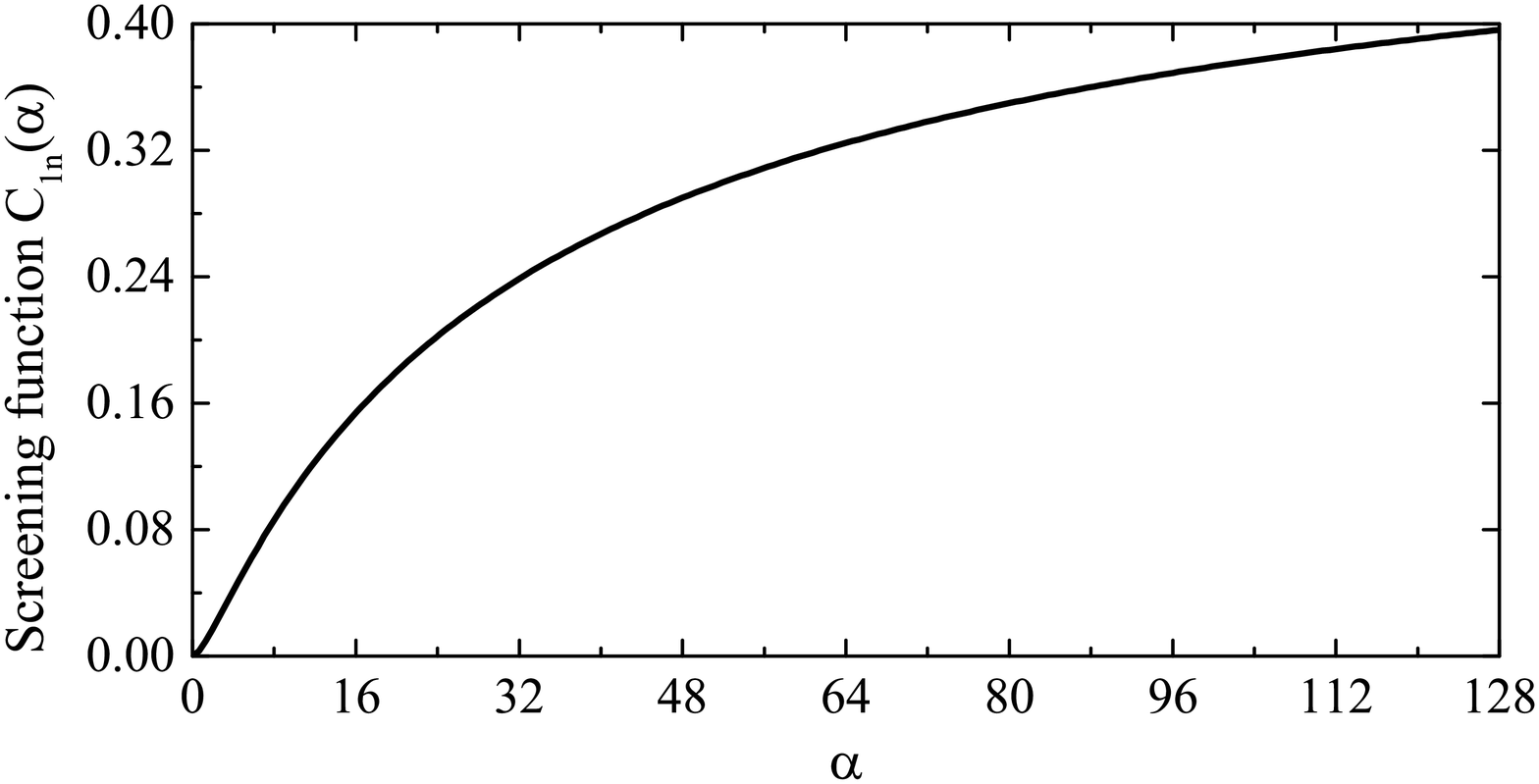}
	\caption{\label{Fig:CexpI}
		Screening function for exponential relaxation {\em via} intra-group processes in the intrinsic nodes.}
\end{figure}

%%%%%%%%%%%%%%%%%%%%%%%%%%%%%%%%
\subsection{Intra-group recombination. Super-exponential relaxation in the extrinsic node group.}

We consider strong pumping of $n$-doped node of WSM, such that the Fermi levels obey the inequalities
\begin{equation}
T\ll |\mu_v|\ll \mu < \mu_c,   
\end{equation}
where $\mu$ is the Fermi level at equilibrium, and the non-equilibrium value is  $\mu_c=\mu+\delta\mu_c$. The pumping is assumed to be strong enough to push the hole quasi-Fermi level into valence band, $\mu_v<0$. Due to the high density of the residual carries $\mu_c$ remains almost unaffected.

Again, it is sufficient to model distribution functions as step-wise. The non-equilibrium densities are given by
\begin{gather}
n=\frac{\eta}{2\pi^2}\frac{\mu^2\delta\mu_c}{v_0^3},\\
p=\frac{\eta}{6\pi^2}\frac{|\mu_v|^3}{v_0^3}.
\end{gather}
The equal excess density condition relates the two quasi-Fermi levels
\begin{equation}
|\mu_v|=\sqrt[3]{3\mu^2\delta\mu_c}
\end{equation}
The dielectric function reads
\begin{equation}
\epsilon(0,q)=1+\frac{2\alpha_{\eta}}{\pi}\frac{\mu^2+\mu_v^2}{\omega_q^2}.
\end{equation}
The statistical factor can be expressed in a closed form
\begin{equation}
\mathcal{S}=C_{2d}(\alpha_{\eta})\frac{\mu^2}{\eta^{2/3}}\frac{p^{2/3}}{\eta^{2/3}}+\tilde{C}_{2d}(\alpha_{\eta})\mu\frac{p}{\eta},
\end{equation}
where the second term corresponds to the next order of expansion in powers of $\mu_v/\mu$. Explicit form of screening functions shown in Fig.~\ref{Fig:Csuper} is
\begin{gather}
C_{2d}(\alpha)=\frac{\alpha^2}{4\pi^{11/3}6^{1/3}}\left[\left(\alpha+\frac{3\pi}{2}\right)\ln{\left(\frac{\pi+2\alpha}{2\alpha}\right)}-\frac{\pi(7\pi+4\alpha)}{4(\pi+2\alpha)}\right]\\
C_{2d}(\alpha)\propto -\frac{6^{2/3}\alpha^2\ln{\alpha}}{16\pi^{8/3}}~(\alpha\rightarrow 0)\\
C_{2d}(\alpha)\rightarrow \frac{6^{2/3}}{144\pi^{2/3}}~(\alpha\rightarrow \infty)
\end{gather}
\begin{gather}
\tilde{C}_{2d}(\alpha)=\frac{\alpha^2}{12\pi^2}\frac{9\pi^2+11\pi\alpha+6\alpha^2}{(\pi+2\alpha)^2}-\frac{\alpha^{3/2}(\pi+\alpha)}{(2\pi)^{5/2}}{\rm arctg}\left(\sqrt{\frac{\pi}{2\alpha}}\right)\\
\tilde{C}_{2d}(\alpha)\propto -\frac{\alpha^{3/2}}{8\sqrt{2\pi}}~(\alpha\rightarrow 0)\\
\tilde{C}_{2d}(\alpha)\rightarrow \frac{1}{15}~(\alpha\rightarrow \infty)
\end{gather}

\begin{figure}[t]
	\includegraphics[width=150mm]{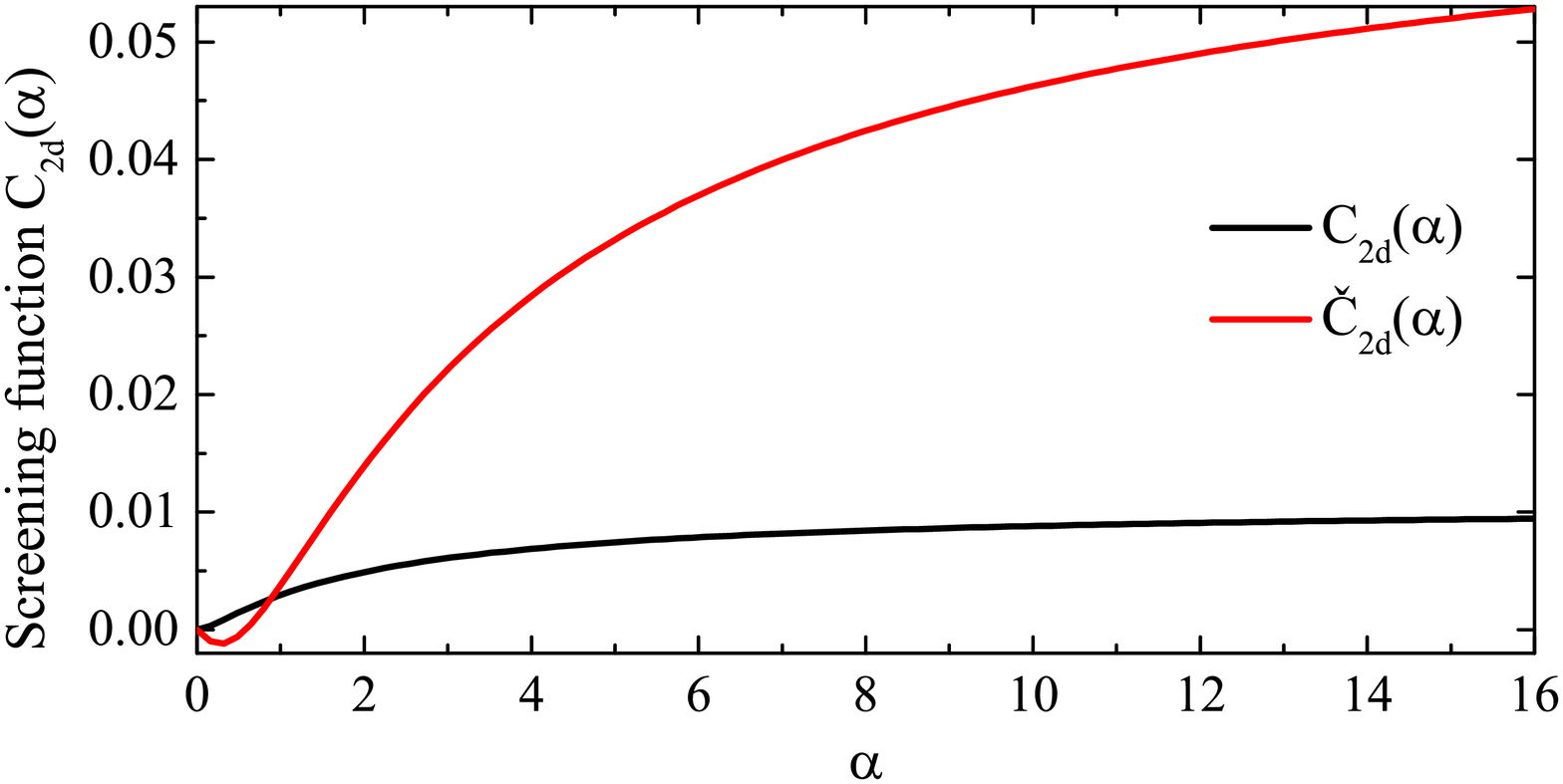}
	\caption{\label{Fig:Csuper}
		Screening function for super-exponential relaxation {\em via} intra-group processes in the extrinsic nodes.}
\end{figure}

%%%%%%%%%%%%%%%%%%%%%%%%%%%%%%
\subsection{Intra-group recombination. Exponential relaxation in the extrinsic node group.}

At the final stages of relaxation in extrinsic $n$-doped nodes, both quasi-Fermi  levels reside in the conduction band, $\mu_c,\mu_v\gg T>0$. The distribution function of conduction electrons can be modelled as step-wise, while the valence band distribution function differs from unity by an exponentially small factor:
\begin{equation}
f_v(E)=1-\exp\left(\frac{E-\mu_v}{T}\right)    
\end{equation}

The non-equilibrium carrier densities are
\begin{gather}
n=\frac{\eta}{2\pi^2}\frac{\mu^2\delta\mu_c}{v_0^3}\\
p\approx\frac{\eta T^2\delta\mu_v}{v_0^3}{\rm e}^{-\frac{\mu}{T}},
\end{gather}
while the Fermi levels are linked by equal excess density condition
\begin{equation}
\delta\mu_c\approx\frac{2T^2}{\mu^2}{\rm e}^{-\frac{\mu}{T}}\delta\mu_v.
\end{equation}
When the number of holes is exponentially small, only majority carriers contribute to screening:
\begin{equation}
\epsilon(0,q)=1+\frac{2\alpha_{\eta}}{\pi}\frac{\mu^2}{\omega_q^2}
\end{equation}
The statistical factor can be expressed in terms of series over $\mu/T$
\begin{equation}
\mathcal{S}=C_{2n}(\alpha_{\eta})\mu\frac{\mu}{T}\frac{p}{\eta}+\tilde{C}_{2n}(\alpha_{\eta})\mu\frac{p}{\eta},
\end{equation}
where the screening functions illustrated in Fig.~\ref{Fig:CexpE} are
\begin{gather}
C_{2n}(\alpha)=\frac{\alpha^2}{12\pi^3}\left[\left(\alpha+\frac{3\pi}{2}\right)\ln{\left(\frac{\pi+2\alpha}{2\alpha}\right)}-\frac{\pi(7\pi+4\alpha)}{4(\pi+2\alpha)}\right]\\
C_{2n}(\alpha)\propto -\frac{\alpha^2\ln{\alpha}}{8\pi^2}~(\alpha\rightarrow 0)\\
C_{2n}(\alpha)\rightarrow \frac{1}{72}~(\alpha\rightarrow \infty),
\end{gather}
and $\tilde{C}_{2n}(\alpha)$ has a very simple form
\begin{equation}
\tilde{C}_{2n}(\alpha)=\frac{5}{24}\frac{\alpha^2}{(\pi+2\alpha)^2}\\
%\tilde{C}_{2n}(\alpha)\propto \frac{5}{24}\frac{\alpha^2}{\pi^2}~(\alpha\rightarrow 0)\\
%\tilde{C}_{2n}(\alpha)\rightarrow \frac{5}{96}~(\alpha\rightarrow \infty)
\end{equation}

\begin{figure}[t]
	\includegraphics[width=150mm]{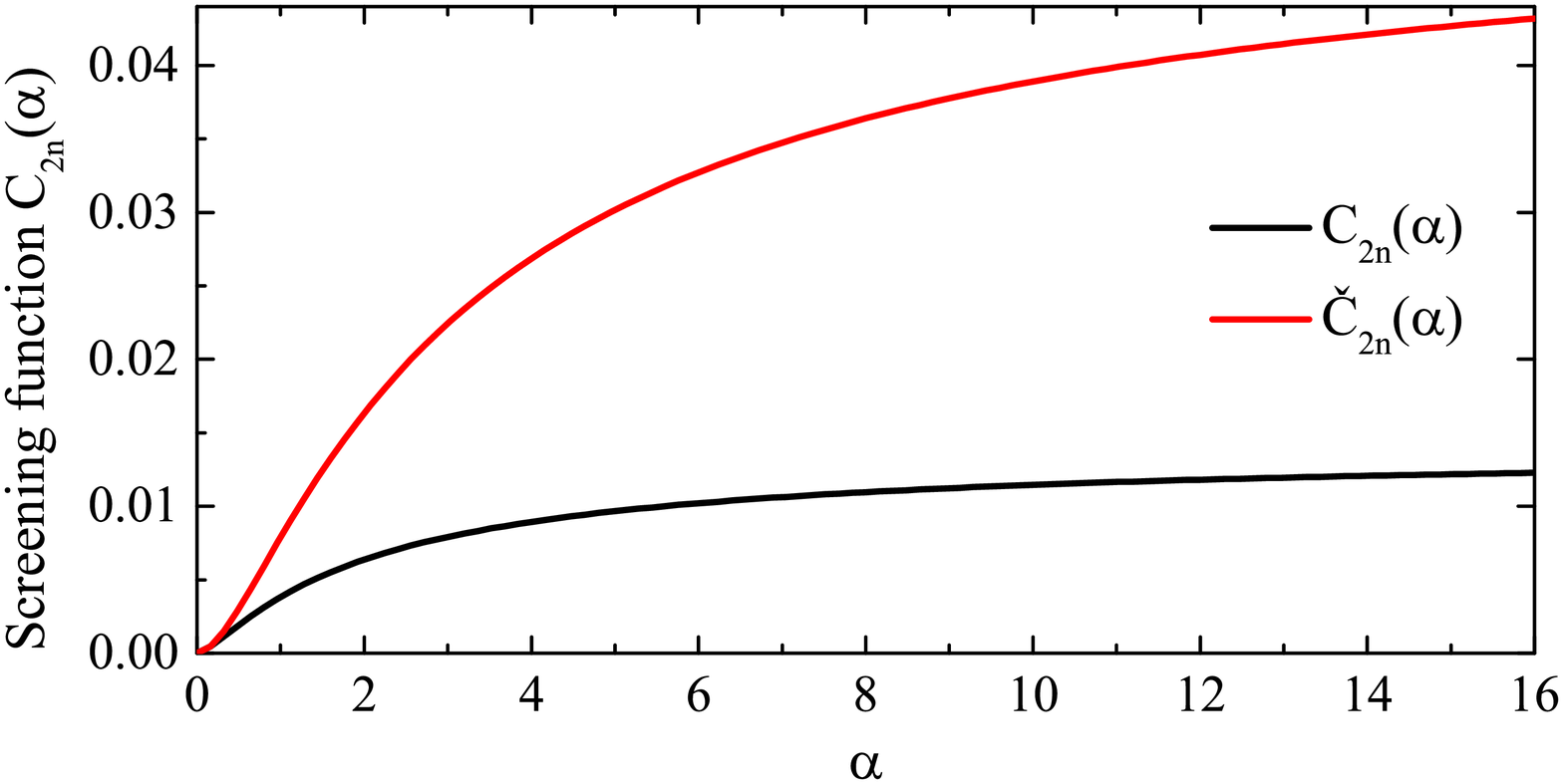}
	\caption{\label{Fig:CexpE}
		Screening function for exponential relaxation {\em via} intra-group processes in the extrinsic nodes.}
\end{figure}

\section{Discussion of the relaxation dynamics in multi-group WSM}

In multi-group WSM the picture of relaxation becomes much more complicated. For two node groups, $W_1$ and $W_2$, there are non-equilibrium densities $p_{1,2}$ appended to the equilibrium ones of rather n-type or p-type, with degenerate, non-degenerate or intermediate statistics, depending on temperature. In addition, there are two relaxation channels, namely intra-group AR and inter-group AR, and the latter leads to interconnection between $p_1(t)$ and $p_2(t)$, as well as screening. Also, there is the intra-band Coulomb or phonon-assisted scattering between the carriers occupying different node groups, which tends to equilibrate quasi-Fermi levels of the groups $\delta\mu_{1,2}$. However, such processes involve large transfer of momentum (equal to distance between pairs of the nodes of $W_1$ and $W_2$), so they are usually slower than the inter-band processes and are neglected here. Depending on details of a particular system and excitation, the shape of relaxation can be quite intricate, and even if relaxation within each node group is exponential, the total dynamics is {\it multi-exponential}.

For the multi-group WSM, considered in the main text, AR processes involving the intraband transitions in $W_1$ dominate due to massive occupation of $W_1$. So, relaxation of $p_2$ is governed by the intra-group AR, while the dominant channel of relaxation of $p_1$ is inter-group AR, which can be locked by the {\em geometry protection}.

In this section, we consider the relaxation dynamics of the population inversion at unprotected group $W_1$. In the weak excitation regime carrier distributions in $W_1$ and the valence band of $W_2$ are non-degenerate, so the relaxation is exponential in both node groups, but with different lifetimes given by 
\begin{equation}
\tau^{-1}=C_{2n}(\alpha_{\eta_2})\mu\frac{\mathcal{G}_1}{\eta_2}\frac{\mu}{T}
\end{equation}
for $W_2$ and
\begin{equation}
\tau^{-1} = C_{4n}T\frac{\mathcal{G}_2}{\eta_2}\frac{T^2}{\mu^2}
\end{equation}
with $C_{4n}\approx2.176$ for $W_1$. Relaxation of the total concentration $p(t)=p_1+p_2$ is {\em multi-exponential} in this case.

For stronger excitation the non-equilibrium carrier distributions become degenerate, so the evolution of $p_2(t)$ is given by
\begin{gather}
p_2(t) =p_2(0) (1-t / 3\tau_{0})^3\\ \tau^{-1}_{0} = C_{2d}(\alpha_{\eta_2}) \frac{\mathcal{G}_1}{\eta_2^{2/3}}\frac{\mu^2}{v_0 [p_2(0)]^{1/3}},
\end{gather}
while
\begin{gather}
p_1(t) = \frac{{p_1}(0)}{1+t/\tau_{0}}, \\
\tau^{-1}_{0} = C_{4d} \frac{\mathcal{G}_2}{\eta_1\eta_2}\frac{v_0^3 p_1(0)}{\mu^2}
\end{gather}
with $C_{4d}=17\pi^2/16\approx10.486$. Numerical solution of Eq.(15) from the main text for unprotected $p_1$ and the corresponding analytical limits are shown in Fig.~\ref{Fig:Dyn_interS}.

\begin{figure}[t]
	\includegraphics[width=150mm]{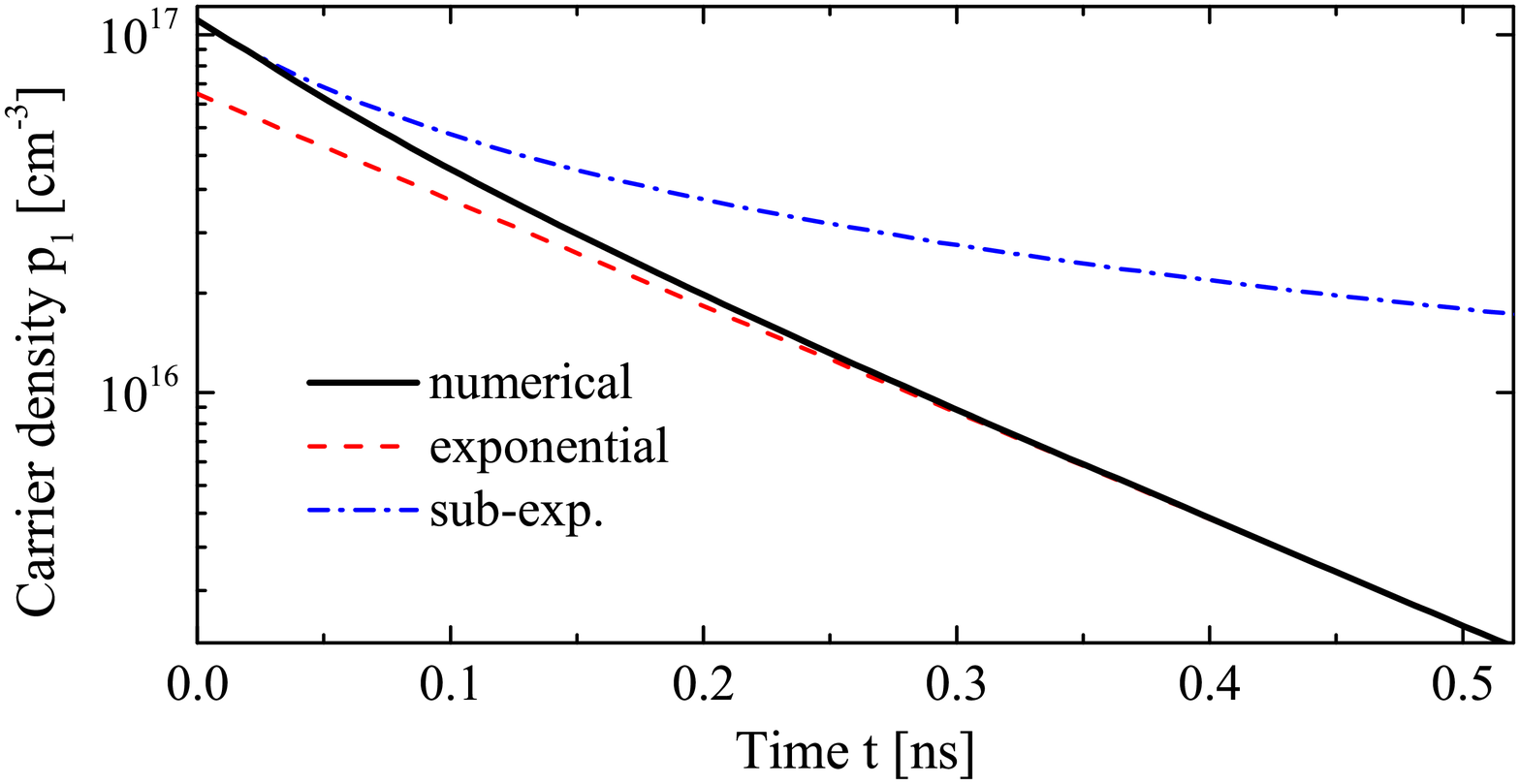}
	\caption{\label{Fig:Dyn_interS}
		Dynamics of the population inversion $p_1(t)$ in WSM with 8-fold degenerate intrinsic node group $W_1$ and 16-fold degenerate extrinsic group $W_2$ without {\em geometrical protection} of $W_1$ ($v_1<v_2$).}
\end{figure}

\end{widetext}

\end{document}